\documentclass[journal]{IEEEtran}
 \usepackage{cite}  

\usepackage{amsmath}
\usepackage{hyperref}
\usepackage{url}
\usepackage{graphicx}  
\usepackage[font=small,labelfont=bf]{caption} 
\usepackage{tikz}
\def\checkmark{\tikz\fill[scale=0.4](0,.35) -- (.25,0) -- (1,.7) -- (.25,.15) -- cycle;} 
\usepackage{float}  
\def\BibTeX{{\rm B\kern-.05em{\sc i\kern-.025em b}\kern-.08em
    T\kern-.1667em\lower.7ex\hbox{E}\kern-.125emX}}
\begin{document}

\title{A Survey on Figure Classification Techniques in Scientific Documents}

\author{
\IEEEauthorblockN{Dhote Anurag Radhesham$^{1}$, Mohammed Javed$^{1}$, David S Doermann$^{2}$}

\IEEEauthorblockA{
\textsuperscript{1}Department of IT, Indian Institute of Information Technology, Allahabad, India \\
\textsuperscript{2}Department of CSE, University at Buffalo, Buffalo, NY, USA \\
\text{Email:\{mit2021082@iiita.ac.in, javed@iiita.ac.in, doermann@buffalo.edu\}}}
}


\maketitle

\begin{abstract}
Figures visually represent an essential piece of information and provide an effective means to communicate scientific facts. Recently there have been many efforts toward extracting data directly from figures, specifically from tables, diagrams, and plots, using different Artificial Intelligence and Machine Learning techniques. This is because removing information from figures could lead to deeper insights into the concepts highlighted in the scientific documents. In this survey paper, we systematically categorize figures into five classes - tables, photos, diagrams, maps, and plots, and subsequently present a critical review of the existing methodologies and data sets that address the problem of figure classification. Finally, we identify the current research gaps and provide possible directions for further research on figure classification.
\end{abstract}

\vspace{5pt}

\section{\textbf{Introduction}}
Classification of images finds tremendous applications in various fields such as automobile, healthcare, agriculture, surveillance, and document analysis \cite{ATM36944, RaniBR2018ClassificationOV, LIU2021223, Kumar2011SurveyOT}. In scientific documents, different graphical visualizations such as tables, photos, diagrams, maps, and plots convey specific facts that are more effective than simple text. This factual information improves comprehension. Hence, extracting underlying information represented by figures is an important task. In general, it is referred to as figure mining. Figure mining includes enhancing the figure design, outlining the data represented by figures, detecting plagiarized documents, etc. The figure mining pipeline consists of (i) figure extraction from academic documents, (ii) classification of figures, and (iii) data extraction from each figure type. This paper aims to survey figure classification techniques and their related datasets comprehensively.  

\begin{figure}[htbp]
\centering
    \includegraphics[scale=0.5]{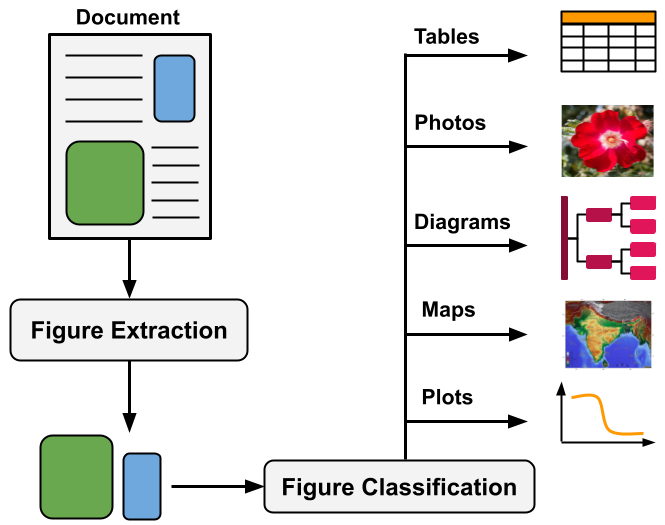}
        \caption{Figure Classification Overview}
        \label{fig:1}
\end{figure}

To address the problem of figure classification, it is crucial to detect and extract the figures from the respective documents using document segmentation techniques, as illustrated in Fig-\ref{fig:1}. Generally, a document image may be segmented into text and non-text components. The non-text details are then further processed to classify them into an appropriate category. Much research has been done on the textual processing of documents. But as far as figures are concerned, there need to be more state-of-the-art methods that classify the scientific figures in their appropriate category. Chart image classification has recently interested many research groups \cite{davila_chart_2021}. This paper aims to highlight the work on chart image classification and include results that include other figure types. The techniques used for classification can be divided into handcrafted-based methods and deep learning-based methods. 

The hand-crafted methods manually extract features using traditional feature extraction techniques, then classify the figures using machine learning models. On the other hand, deep learning techniques automatically learn features and classify the figures. Various approaches employed in these two categories are discussed in detail in the upcoming sections. This follows a discussion on several data sets reported in the related literature.  
The rest of the paper is organized as follows. Section 2 provides information on the existing literature on the figure classification problem, and a summary of significant contributions is shown in Table\ref{table:1}. Section 3 includes a discussion of datasets used in recent works, and details of a few publicly available datasets are summarised in Table-\ref{table:2}. Section 4 provides pointers for future research work and many interesting problems that still need to be addressed in figure classification.\\

\section{\textbf{Overview of Figure Classification Problem}}

\begin{table*}[htbp] 
\normalsize
\centering
\caption{Figure types considered for classification in the existing literature}
\begin{tabular}{|l |c| c| c| c| c |c| c|}
\hline
\textbf{Literature} & \textbf{Table} & \textbf{Plot} & \textbf{Diagram} & \textbf{Photo} & \textbf{Equation} & \textbf{Geometric Shape} & \textbf{Map}\\ [0.5ex]
\hline
\hline 
Siegel et al. \cite{siegel_figureseer_2016} & \checkmark & \checkmark & \checkmark & & & & \\
\hline
Lee et al.\cite{lee_viziometrics_2018} & \checkmark & & & & \checkmark & & \\
\hline
Morris et al.\cite{morris_slideimages_2020} & \checkmark & \checkmark & \checkmark & \checkmark & & & \checkmark \\
\hline
Jobin et al.\cite{kv_docfigure_2019} & \checkmark & \checkmark & \checkmark & \checkmark & & \checkmark & \checkmark \\
\hline
Andrearczyk and Müller\cite{andrearczyk_deep_2018} & \checkmark & \checkmark & \checkmark & \checkmark & & & \\
\hline
Almakky et al.\cite{almakky_stacked_2018} & \checkmark & \checkmark & \checkmark & \checkmark & & & \\
\hline
Lu et al.\cite{lu_automated_2009} & & \checkmark & \checkmark & & \checkmark & & \\
\hline
Cheng et al.\cite{cheng_graphical_2013} & & & \checkmark & \checkmark & & & \\
\hline
Lagopoulos et al.\cite{lagopoulos_classifying_2019} & & & \checkmark & \checkmark & & & \\
\hline
Giannakopoulous et al.\cite{giannakopoulos_visual-based_2015} & & & & & & \checkmark & \checkmark\\
\hline
\end{tabular}
\label{table:1}
\end{table*}

Figures are visualizations used in scientific literature to convey information and enhance comprehension. Figures often represent data that would otherwise be difficult to process if conveyed by the text. Figures are commonly categorized into well-known classes, such as tables, plots, diagrams, photos, equations, geometric shapes, maps, etc. Classes considered under the classification of figures can vary widely depending on the research field\cite{lee_viziometrics_2018}. Giannakopoulos et al. \cite{giannakopoulos_visual-based_2015} identify charts, diagrams, geometric shapes, maps, and photographs as the classes for the figure classification problem. Lee et al. \cite{lee_viziometrics_2018} also considered the table a separate figure class in addition to plots, diagrams, photos, and equations. Table--\ref{table:1} summarizes the different figure types present in the existing literature. 
It can be observed from the table that the figure categories like tables, plots, diagrams, and photos are popular figure types as compared to equations, geometric shapes, and maps. Considering the previous taxonomies, this paper's figures are divided into Tables, Photos, Diagrams, Plots, and maps. These five categories cover all the existing categories explored so far. 
\subsection{\textbf{Table}}
A table is a structure with cells containing text and numeric data. Tables are very efficient at summarizing the textual information between methods that address similar problems. Tables in literature are used for tasks such as comparing existing methods, summarizing the data sets, highlighting observations, etc. Tables are hence recognized as an essential figure type in literature. Table detection and recognition problems have been extensively studied in previous years, and Hashmi et al.\cite{hashmi_current_2021} summarize the existing work in the review. But classifying tables from other figure types remains open to research problems. Only some studies that have included tables while classifying figures use traditional and deep learning approaches. Lee et al.\cite{lee_viziometrics_2018} use bags of visual words to classify tables among other figures such as Photos, Diagrams, Plots, etc. 

Similarly, Jobin et al.\cite{kv_docfigure_2019}, Morris et al.\cite{morris_slideimages_2020}, Siegel et al.\cite{siegel_figureseer_2016} use deep learning techniques to classify tables among other types of figures. Tables are rarely further divided into subcategories as the information conveyed through different table structures does not add to any further comprehension. So there are no subcategories under the class table.

\subsection{\textbf{Photo}}
A photo is generated when light falls on a photosensitive surface like a photographic sensor. Natural and medical images (diagnostic and radiological photos) are considered under the class photos. Depending on the scientific field, the presence of photos varies drastically. Photos are used in literature to provide deep insights on a specific topic, which are difficult to provide using text or other figure types. Jobin et al. [add ref] identified natural and medical images as figure categories in the DocFigure data set. They used a combination of FC-CNN and FV-CNN to classify these figure types. Medical images, commonly used in medical journals, papers, and articles, are further sub-categorized into diagnostic and nondiagnostic images in ImageCLEF2013 and 2016 datasets. Lagopoulos et al.\cite{lagopoulos_classifying_2019}, Almakky et al.\cite{almakky_stacked_2018}, Andrearczyk, and Muller\cite{andrearczyk_deep_2018} consider the ImageCLEF2016 data sets to perform the figure classification task.  

\subsection{\textbf{Diagram}}
A diagram represents the relationship between various parts of a concept. Figures like flowcharts, Gantt charts, schematics, conceptual diagrams, and tree diagrams are considered under the class diagrams. Diagrams improve perception by visualizing the structure and flow of a concept. Therefore, they are ubiquitous  the  scientific literature. Classification of diagrams into their subcategories has yet to be addressed in the literature. However, the existing literature has discussed the problem of the classification of diagrams among other figure types. Jobin et al.\cite{kv_docfigure_2019} considered flow charts and tree diagrams as figure types in the classification of figures. Lee et al.\cite{lee_viziometrics_2018} identify diagrams as a crucial figure type and address its classification among other figure types. The bag-of-visual-words-based method is used to classify diagrams from different figure types.

\begin{table*}[htbp] 
\normalsize
\centering
\label{table:2}
\caption{Summary of existing figure datasets and related methods}
\begin{tabular}{|l | c | l| c| c| l |c|}
\hline
\textbf{Authors} & \textbf{Year} & \textbf{Dataset} & \textbf{\#Samples} & \textbf{\#Categories} & \textbf{Model} & \textbf{Accuracy} \\
\hline
\hline
Savva et al.\cite{savva_revision_2011} & 2011 & Self-acquired & 2601 & 10 & SVM & 96.00\%\  \\ 
\hline
Gao et al.\cite{gao_view_2012} & 2012 & Self-acquired & 300 & 3 & SVM & 97.00\%\  \\
\hline
Kartikeyani &  & &   &  & MLP & 69.68\%\ \\ 
and & 2012 & Self-acquired & 155 & 8 & KNN & 78.06\%\ \\
Nagarajan\cite{karthikeyani_machine_2012} & & & & & SVM & 76.77\%\ \\ 
\hline
Cheng et al.\cite{cheng_graphical_2013} & 2014 & Self-acquired & 1707 & 3 & MLP & 96.1\%\\
\hline
Liu et al.\cite{liu_chart_2015} & 2015 & DeepChart & 5000 & 5 & CNN + DBN & 75.40\%\ \\ 
\hline
Siegel et al.\cite{siegel_figureseer_2016} & 2016 & FigureSeer & 60000 & 7 & AlexNet & 84\%\ \\ 
& & & & & ResNet-50 & 86\%\ \\
\hline
Amara et al.\cite{amara_convolutional_2017} & 2017 & Self-acquired & 3377 & 11 & CNN & 89.50\%\\
\hline
Jung et al.\cite{jung_chartsense_2017} & 2017 &
Chart-Sense & 6997 & 10 & GoogleNet & 91.30\%\ \\
\hline
Almakky et al.\cite{almakky_stacked_2018} & 2018 & ImageCLEF\cite{sanderson2019imageclef}
 & 10942 & 30 & SDAE + SVM & 64.30\%\ \\
\hline
Balaji et al.\cite{balaji_chart-text_2018} & 2018 & Self-acquired
 & 6000 & 2 & CNN & 99.72\%\ \\
\hline
Chagas et al.\cite{chagas_evaluation_2018} & 2018 & Chart-Vega & 14471 & 10 & ResNet-50 & 76.76\%\ \\ 
& & & & & Inception-V3 & 76.77\%\ \\
\hline
Dai et al.\cite{dai_chart_2018} & 2018 & Self-acquired & 11,174 & 5 & ResNet & 98.89\%\ \\ 
& & & & & GoogLeNet & 99.07\%\ \\
& & & & & AlexNet & 99.48\%\ \\
& & & & & VGG-16 & 99.55\%\ \\
\hline
Liu et al.\cite{liu_data_2019} & 2019 & Self-acquired & 2500 & 2 & VGG-16 & 96.35\%\ \\ 
\hline
Davila et al.\cite{davila_icdar_2019} & 2019 & Synthetic & 202550 & 7 & ResNet-101 & 99.81\%\ \\ 
\hline
Jobin et al.\cite{kv_docfigure_2019} & 2019 & Doc-Figure & 33000 & 28 & FC-CNN + FV-CNN & 91.30\%\ \\ 
\hline
Bajic et al.\cite{bajic_data_2020} & 2020 & Self-acquired & 2702 & 10 & VGG-16 & 89.00\%\ \\ 
\hline
Araujo et al.\cite{araujo_real-world_2020} & 2020 & Self-acquired & 21099 & 13 & Xception & 95.00\%\ \\
\hline
Morris et al.\cite{morris_slideimages_2020} & 2020 & SlideImages & 3629 & 9 & MobileNetV2 & 80.00\%\ \\ 
\hline
Luo et al.\cite{luo_chartocr_2021} & 2021 & Chart-OCR & 386966 & 3 & CNN & 93.30\%\ \\ 
\hline
Davila et al.\cite{davila_icpr_2021} & 2021 & UB-PMC & 22924 & 15 & DenseNet-121 + ResNet-152 & 92.80\%\ \\ 
\hline
\end{tabular}
\end{table*}

\subsection{\textbf{Map}}
A map is a symbolic representation of the characteristics of a place or distribution. The map includes subcategories such as Geographical maps, Scientific maps, TreeMaps, and other geographical representations. Maps are used to describe various features localized in a particular area. Using scientific maps could lead to new insights into existing communities, concepts, and demographics based on map type. Hence it is essential to include maps as a  figure type. Many researchers do not consider maps when addressing figure classification tasks. Giannakopoulos et al.\cite{giannakopoulos_visual-based_2015}, Jobin et al.\cite{kv_docfigure_2019}, Morris et al.\cite{morris_slideimages_2020} include several types of maps in the dataset. Jobin et al. have incorporated Treemaps and Geographical maps into the DocFigure dataset. At the same time, Morris et al. include only geographical maps. As far as the author knows, scientific maps are not included in the existing literature.

\subsection{\textbf{Plot}}
The plot is a visual technique representing the relationships between two or more variables.  Plots are widely used in the scientific literature to convey results with more clarity. There are various subcategories of plots, such as scatter, bar, pie, line, area, etc. Plots have strong representative power and simple rules and have been used in multiple research fields; hence they are considered significant figure types. As plots can be divided into various subcategories, which are also widely used in scientific literature, it has been addressed in existing works more than the other figure types. The following subsections have discussed a few traditional and deep learning approaches for addressing chart image classification. 

\section{\textbf{Related Work}}
The approaches implemented in the present work can be divided into traditional and deep learning categories. The figure classification problem has been addressed more in the bio-medical field than in other areas. This could be because a state-of-the-art data set is designed for automated figures analysis in the biomedical literature called ImageCLEF\cite{sanderson2019imageclef}. A detailed discussion regarding various approaches used for figure classification is provided in the following sub-sections. In addition to this, specifically, chart classification techniques are also summarized in detail. 
\subsubsection{\textbf{Traditional Approaches}}
Traditional approaches rely on feature extraction methods used in computer vision. Features are manually extracted from the figures and then represented in mathematical form for further processing. These mathematical representations act as input to the classifiers, following the traditional method-based approach Savva et al.\cite{savva_revision_2011} present a system that automatically remodels the visualizations to increase visual comprehension. The authors use low-level image features for classification, and to improve further classification, they use text-level features. The performance is tested by training a multiclass SVM classifier on a corpus containing 2601 chart images labeled with ten categories, following Gao et al.'s manual extraction path.\cite{gao_view_2012}, propose VIEW, a system that automatically extracts information from raster-format charts. The authors separate the textual and graphical components and classify the given chart image based on the graphic elements extracted from the visual components using SVM. 

The text is limited to three chart categories of bar charts, pie charts, and line graphs, with 100 images for each category collected from various real-world digital resources. Instead of taking an image as input, Karthikeyani and Nagarajan\cite{karthikeyani_machine_2012} present a system to recognize chart images from PDF documents using eleven texture features that are part of the Gray Level Co-Occurrence Matrix. A chart image is located in the PDF Document database, and the features are extracted and fed to the learning model. SVM, KNN, and MLP are the classifiers used for getting the classification results. Cheng et al.\cite{cheng_graphical_2013} employ a multimodal approach that uses text and image features. These features are provided as input to MLP. The output is characterized as fuzzy sets to get the final result. The corpus contains 1707 figures with three categories and a 96.1\%\ classification result. ReVision pioneered the technique for chart image classification and would act as a state-of-the-art method for future methods.

\subsubsection{\textbf{Deep Learning Approaches}}
Liu et al.\cite{liu_chart_2015} used a combination of Convolutional Neural Networks(CNN) and Deep Belief Networks (DBN) to capture high-level information present in deep hidden layers; fully Connected Layers of Deep CNN are used to extract deep hidden features. DBN is then used to predict the image class on the mentioned deep hidden features. Authors use the transfer learning concept and then perform fine-tuning to prevent overfitting. The data set included more than $5,000$ images of charts in the categories of pie charts, scatter charts, line charts, bar charts, and flow charts. Deep features are useful over primitive features to provide better stability and scalability to the proposed framework.

Given the results of CNN in the classification of natural images, Siegel et al.\cite{siegel_figureseer_2016} use two CNN-based architectures for figure classification. They evaluate AlexNet and ResNet-50, which are pre-trained on the ImageNet data set and then fine-tuned for figure classification. This transfer learning approach would be prevalent in subsequent works addressing this problem. The proposed frameworks outperformed the state-of-the-art model, ReVision, by a significant margin. ResNet-50 achieved the best classification accuracy of 86\%\ performed on a dataset containing over 60000 images spread across seven categories.  

Amara et al.\cite{amara_convolutional_2017} proposed a CNN-based LeNet model to classify their corpus of 3377 images into 11 categories. The model comprises eight layers: an output layer, one fully connected layer, five hidden layers, and an input layer. The fully connected layer is used as a classifier, while the hidden layers are convolution and pooling layers designed to extract features automatically. A fully connected layer employs softmax activation to classify images into predefined classes. For evaluation of the model's performance, an 80-20 split is performed on the data set for training and assessment. The proposed model performs better than the LeNet and pretrained LeNet architectures with an accuracy of 89.5\%. 

Jung et al. \cite{jung_chartsense_2017} present a classification method using the Caffe deep learning framework and evaluate its efficacy by comparing it with ReVision (a state-of-the-art chart-type classification system). The authors use GoogLeNet for classification and compare its results with shallower networks like LeNet-1 and AlexNet. Fivefold cross-validation is used for calculating the accuracy of the image corpus with 737 - 901 images for each chart type. The text concludes that ChartSense provides a higher classification accuracy for all types of graphs than ReVision.

\begin{table*}[htbp] 
\centering
\normalsize
\caption{Size of the available datasets with different figure types}
\label{table:2}
\begin{tabular}{|l |c| c |c |c| c|}
\hline
\textbf{Figure} & \textbf{UB-PMC \cite{davila_icdar_2019}} & \textbf{Doc-Figure \cite{kv_docfigure_2019}} & \textbf{Chart-Sense \cite{jung_chartsense_2017}} & \textbf{Chart-OCR \cite{luo_chartocr_2021}} & \textbf{Chart-Vega \cite{chagas_evaluation_2018}} \\ 
 \hline
 \hline
Arc & - & - & - & - & $1440$  \\ 

Area &  $172$ & $318$ & $509$ & - & $1440$  \\ 

Block & - & $1024$ & - & - & - \\ 

Bubble & - & $339$ & - & - & -\\ 

Flowchart & - & $1074$ & - & - & -\\ 

Heatmap & $197$ & $1073$ & - & - & -\\ 

Horizontal Bar &   $788$ & - & - & - & -\\ 

Horizontal Interval &   $156$ & - & - & - & - \\ 

Line & $10556$ & $9022$ & $619$ & $122890$ & $1440$ \\ 

Manhattan &   $176$ & - & - & - & - \\ 

Map &   $533$ & $1078$ & $567$ & - & -\\ 

Parallel Coordinate & - & - & - & - & $1339$ \\ 

Pareto & - & $311$ & $391$ & - & -\\ 

Pie &   $242$ & $440$ & $568$ & $76922$ & $1440$\\ 

Polar & - & $338$ & - & - & -\\ 

Radar & - & $309$ & $465$ & - & -\\ 

Re-orderable Matrix & - & - & - & - & $1440$\\ 

Scatter & $1350$ & $1138$ & $696$ & & $1640$\\ 

Scatter-Line &   $1818$ & - & - & - & - \\ 

Sunburst & - & - & - & - & $1440$\\ 

Surface & $155$ & $395$ & - & - & -\\ 

Table & - & $1899$ & $594$ & - & -\\ 

Treemap & - & - & - & - & $1440$\\ 

Venn &   $75$ & $889$ & $693$ & - & -\\ 

Vertical Bar & $5454$ & $1196$ & $557$ & $187154$ & $1512$\\ 

Vertical Box &   $763$ & $605$ & - & - & -\\ 

Vertical Interval &   $489$ & - & - & - & -\\ 
\hline

Total & $22924$ & $33071$ & $5659$ & $386966$ & $14471$ \\ \hline

\end{tabular}

\end{table*}

Almakky et al.\cite{almakky_stacked_2018} developed a stack-auto encoder model for figure classification. They work with the ImageCLEF 2016\cite{sanderson2019imageclef} data set for biomedical subfigures having 30 classes and 10942 images. The data imbalance related to biomedical images has led the authors to use the proposed model. Five autoencoders were trained separately to extract the features in an unsupervised manner. This model is further fine-tuned to retain cohesion using the same binary cross-entropy criterion used to train SDAE layers. An overall accuracy of 64.3\%\ was achieved using the proposed method. Poor overall accuracy compared to other works under the ImageCLEF challenge is attributed to low training samples and the nature of the data set.

With studies adapting the deep learning approach for chart image classification, a comparative study of traditional vs. CNN architectures was required. Chagas et al.\cite{chagas_evaluation_2018} provide a comparative analysis of conventional vs. CNN techniques. The authors evaluate CNN architectures (VGG19, Resnet-50, and Inception-V3) for chart image classification for ten classes of charts. The performance is compared with conventional machine learning approaches such as classifiers Naive Bayes; HOG features combined with KNN, Support Vector Machine, and Random Forest. Pre-trained CNN models with fine-tuned last convolutional layers were used. The authors concluded that the CNN models surpass traditional methods with an accuracy of 77.76\%(Resnet-50) and 76.77\%(Inception-V3) compared to 45.03\%(HOG+SVM).

Limitation in the figure data set was a significant problem in chart mining as both size and categories limited existing datasets. So, Jobin et al.\cite{kv_docfigure_2019} presented DocFigure, a figure classification data set with $33,000$ figures for 28 different categories. To classify figures, the author proposes techniques that utilize the deep feature, deep texture feature, and a combination of both. Among these baseline classification techniques, the authors observed that combining deep feature and deep texture feature classifies images more efficiently than individual feature techniques. The average classification accuracy improved by 3.94\% and 2.10\% by concatenating FC-CNN and FV-CNN over individual use of FC-CNN and FV-CNN, respectively. The overall accuracy of the combined feature methods turned out to be 92.90\%.

Due to the need for benchmarks in the chart mining process, Davila et al.\cite{davila_icdar_2019} summarized the works of different participants in the first edition of the competition on Harvesting Raw Tables from Infographics, which provided data and tools to the chart recognition community. Two data sets were provided for the classification task. One was a synthetically generated AdobeSynth dataset, and the other UB-PMC data set was gathered from the PubMedCentral open-access library. The highest accuracy achieved for the synthetic data set was 99.81\%\ whereas for the PMC data set, it was 88.29\%. In the second edition of the competition, as the PMC set was improved and included in the training phase, the accuracy of models over the PMC set improved significantly to 92.8.\% 

Luo et al. proposed a unified method to handle various chart styles.\cite{luo_chartocr_2021} where they prove that generalization ability can be obtained in deep learning frameworks with rule-based methods. The experiments were carried out on three datasets with more than $300,000$ images with three categories of graphs. In addition to the framework, an evaluation metric for bar, line, and pie charts is also introduced. The authors concluded that the proposed framework performs better than traditional, rule-based deep learning methods. Amara et al.\cite{amara_convolutional_2017} propose a deep learning-based framework that automates the feature extraction step, an improved LeNet convolutional neural network architecture version. Over $90,000$ images of charts from 11 different categories were chosen for the experiments, and the proposed framework performs significantly better than model-based approaches.

\section{Datasets}
There need to be more datasets that contain all the figure types discussed before. DocFigure\cite{kv_docfigure_2019} is one data set that includes tables, flowcharts, and other plots in a combined data set of $33,000$ images. Morris et al.\cite{morris_slideimages_2020} propose SlideImages which includes 9 different classes with $3,629$ images of various figures. Given the popularity of table recognition problems, data sets dedicated to images of tables have been developed over the past decade. Current works employ augmentation methods to cope with the problem of a small data set\cite{davila_chart_2021}.

There has been a significant improvement in size for chart image classification. Revision\cite{savva_revision_2011} dataset, which would be used for further studies for comparison, had only $2,601$ images. The data sets proposed in recent years have more than $20,000$ images. However, the data sets used for classification purposes mainly contain synthetic images. All data sets include the actual chart image in JPG, PNG, or JPEG format and the corresponding annotations in JSON and XML format. These studies ignore 3D charts, hand sketches, and composite figures. There need to be more authentic figure images extracted from documents that do not follow the fixed constraint prevalent in training image samples of existing data sets. Table-\ref{table:3} below shows the types of figures and their corresponding sample sizes. The data sets mentioned in the table are publicly available and were considered in the works of literature mentioned above. 

\begin{figure}[t] 
\scriptsize
\centering
\begin{tabular}{c c c c} 

\includegraphics[width=1.5 cm, height=1.5 cm]{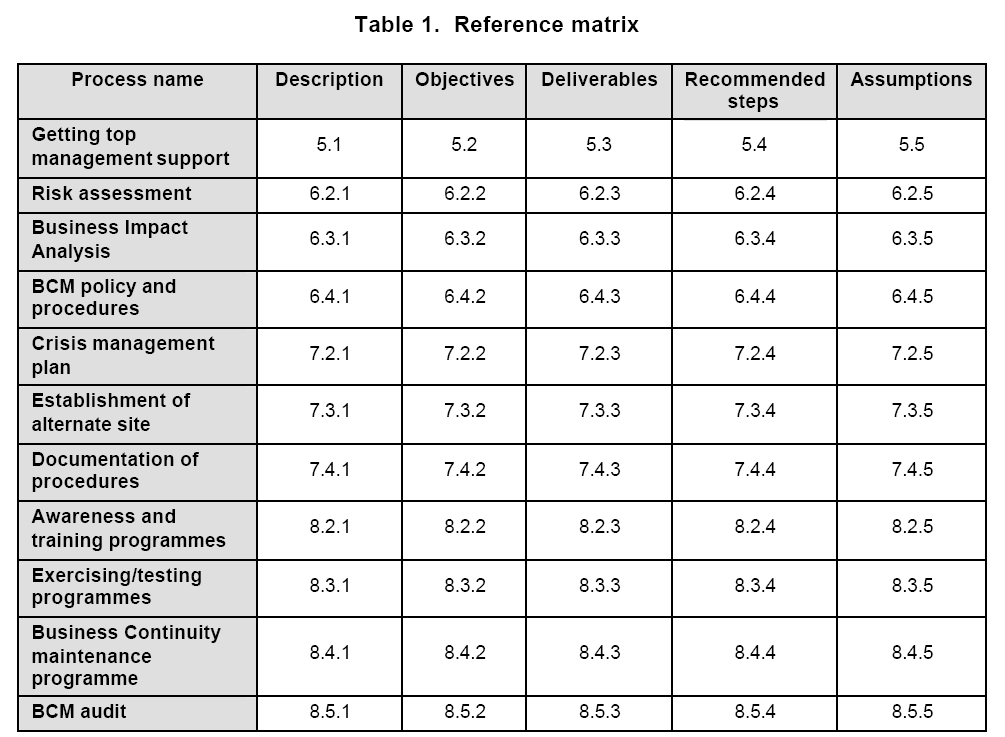} & 
\includegraphics[width=1.5 cm, height= 1.5 cm]{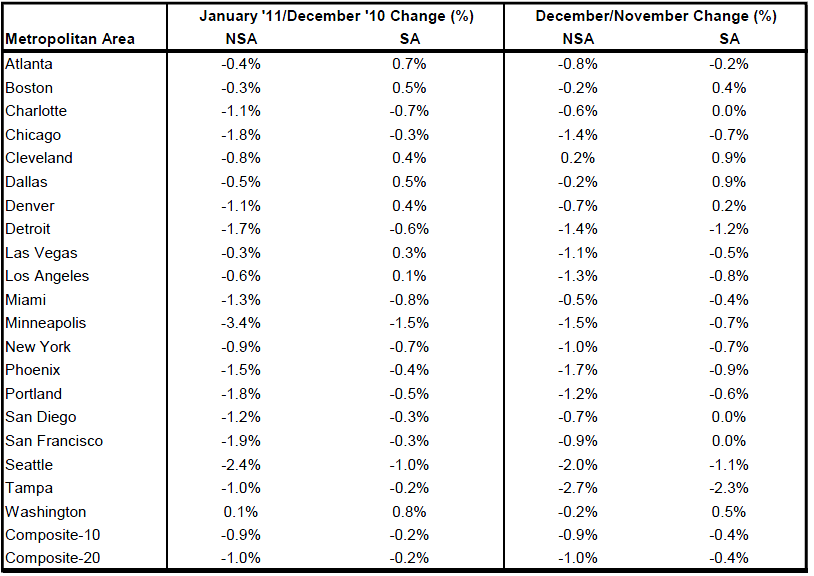} &
\includegraphics[width=1.5 cm, height=1.5 cm]{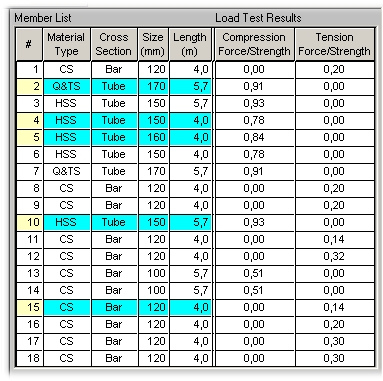} &
\includegraphics[width=1.5 cm, height=1.5 cm]{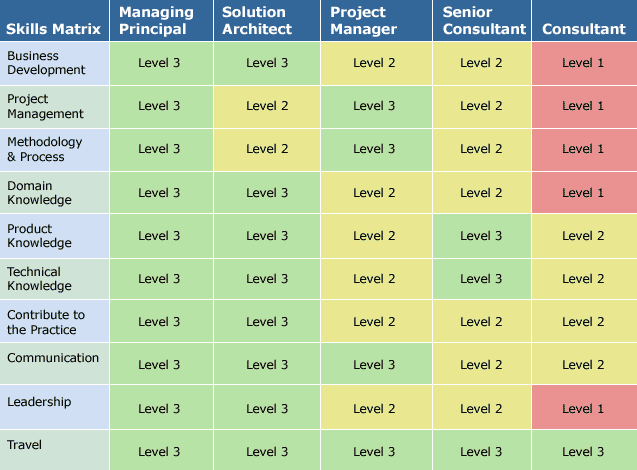}
\\

\text{a. Table - 1}  & \text{b. Table - 2} & \text{c. Table - 3} & \text{d. Table - 4} 
\\
\\

\includegraphics[width=1.5 cm, height=1.5 cm]{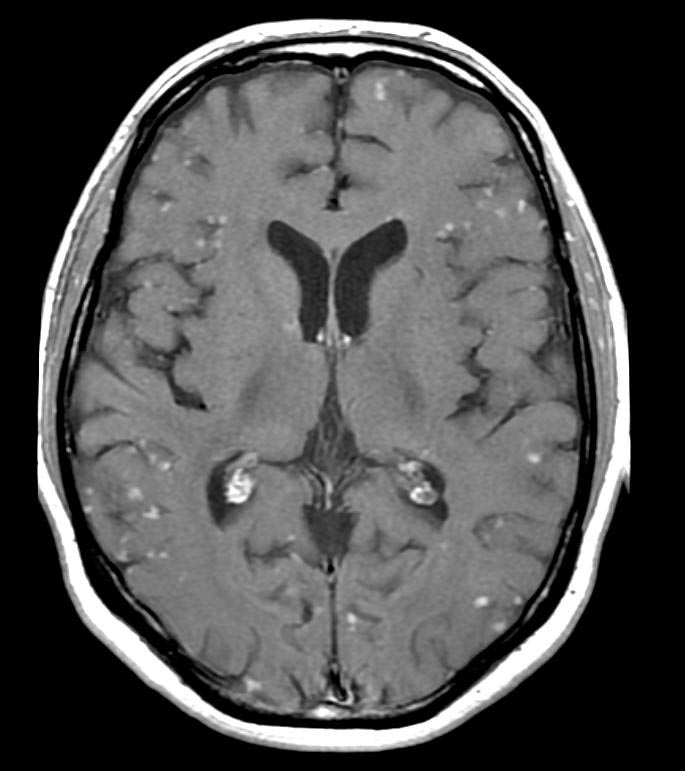} & 
\includegraphics[width=1.5 cm, height=1.5 cm]{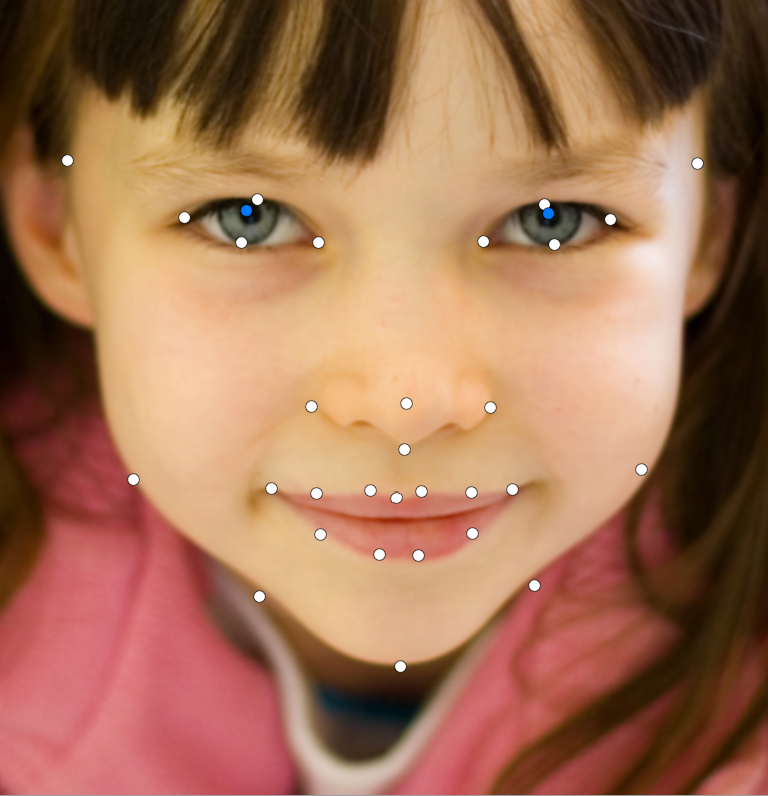} &
\includegraphics[width=1.5 cm, height=1.5 cm]{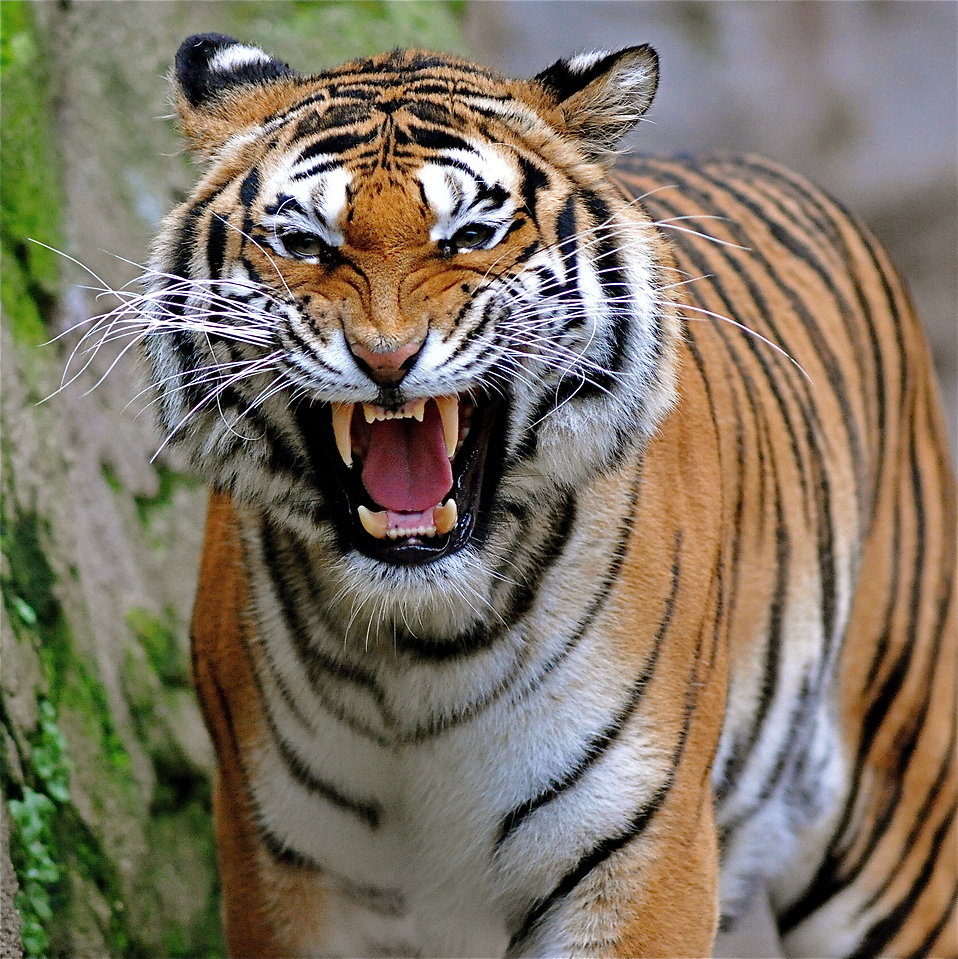} &
\includegraphics[width=1.5 cm, height=1.5 cm]{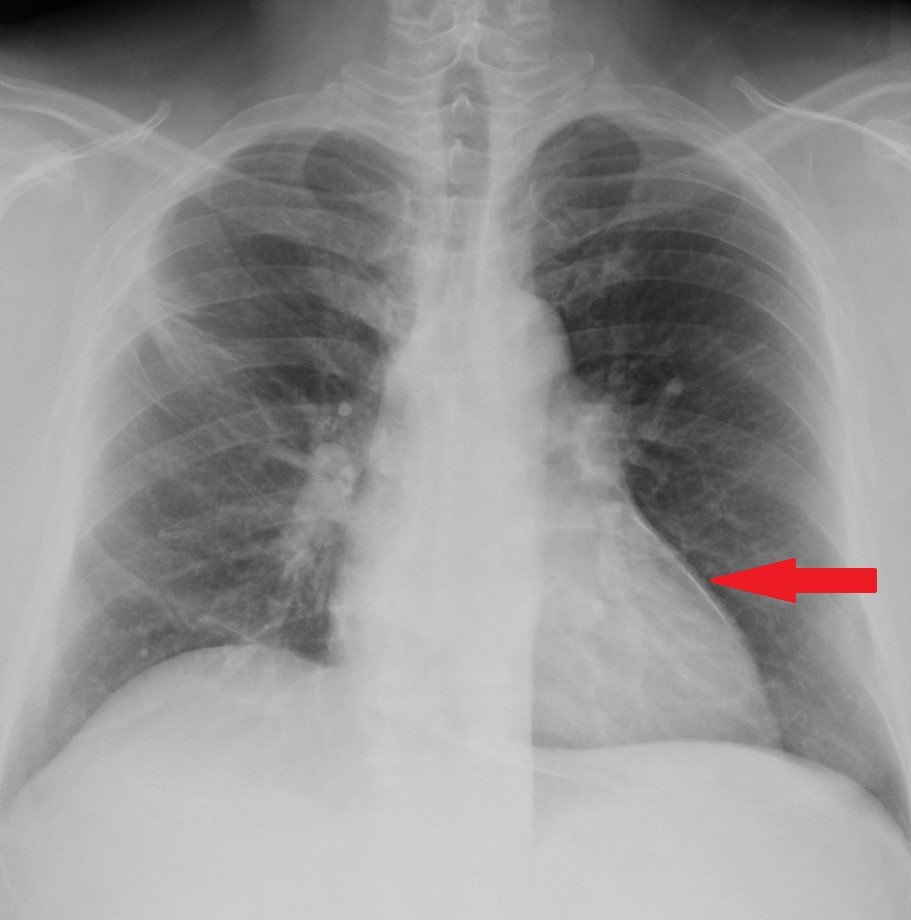}
\\

\text{e. Photo - 1}  & \text{f. Photo - 2} & \text{g. Photo - 3} & \text{h. Photo - 4} 
\\
\\

\includegraphics[width=1.5 cm, height=1.5 cm]{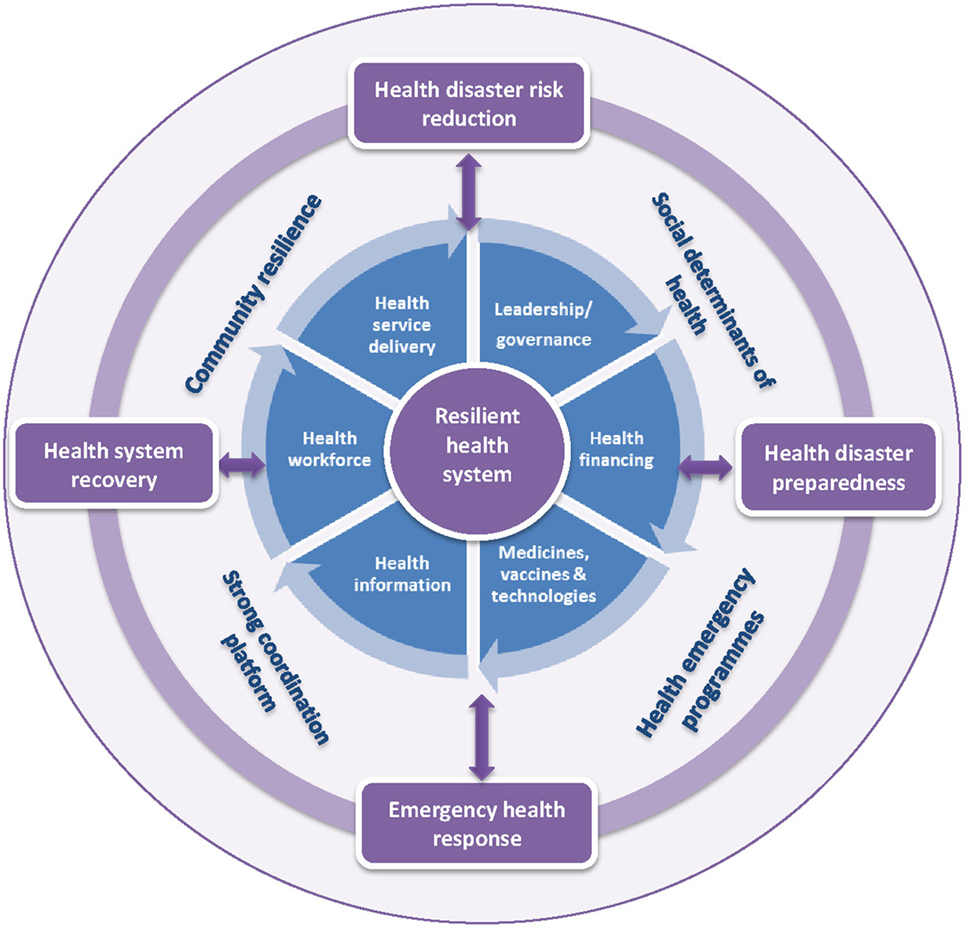} & 
\includegraphics[width=1.5 cm, height=1.5 cm]{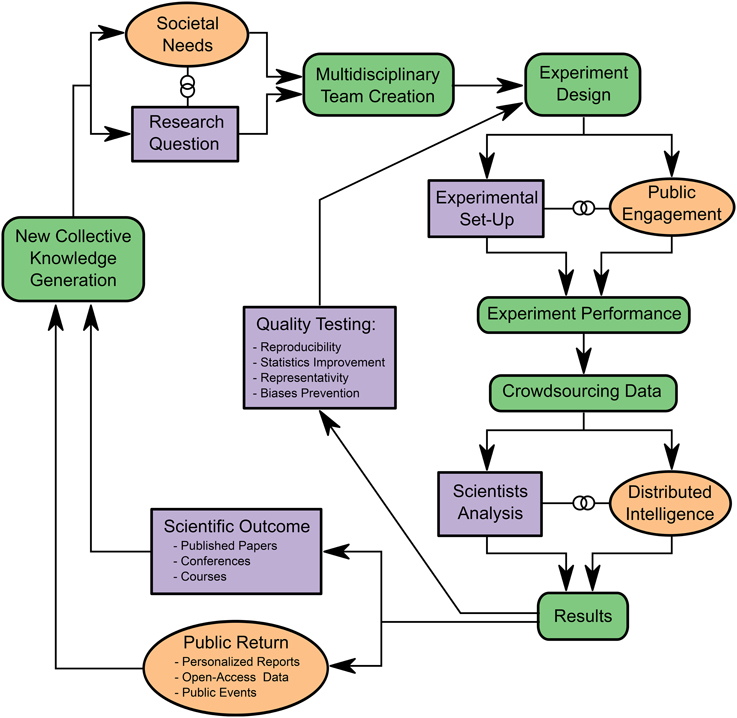} &
\includegraphics[width=1.5 cm, height=1.5 cm]{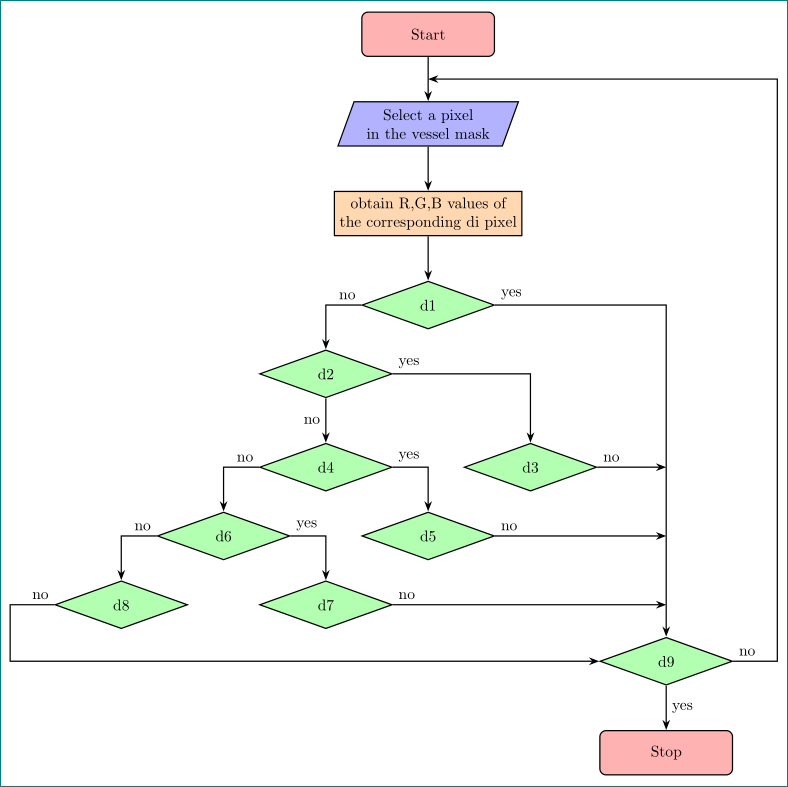} &
\includegraphics[width=1.5 cm, height=1.5 cm]{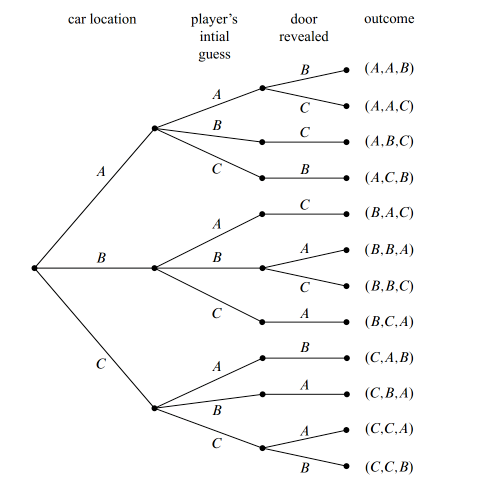}
\\

\text{i. Diagram - 1}  & \text{j. Diagram  - 2} & \text{k. Diagram  - 3} & \text{l. Diagram  - 4} 
\\
\\

\includegraphics[width=1.5 cm, height=1.5 cm]{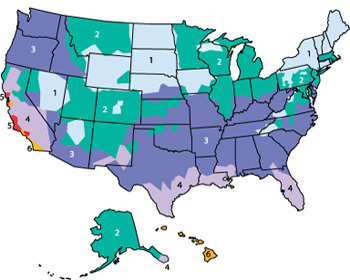} & 
\includegraphics[width=1.5 cm, height=1.5 cm]{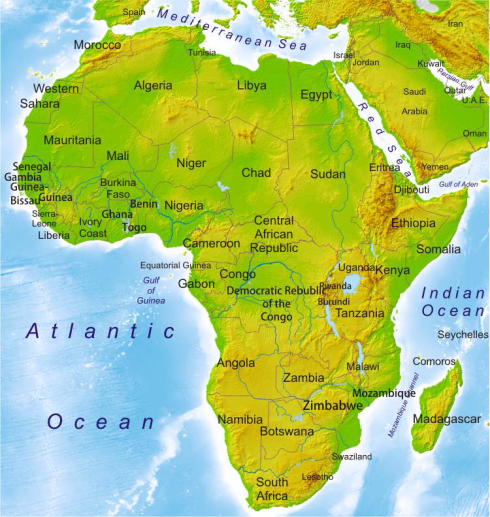} &
\includegraphics[width=1.5 cm, height=1.5 cm]{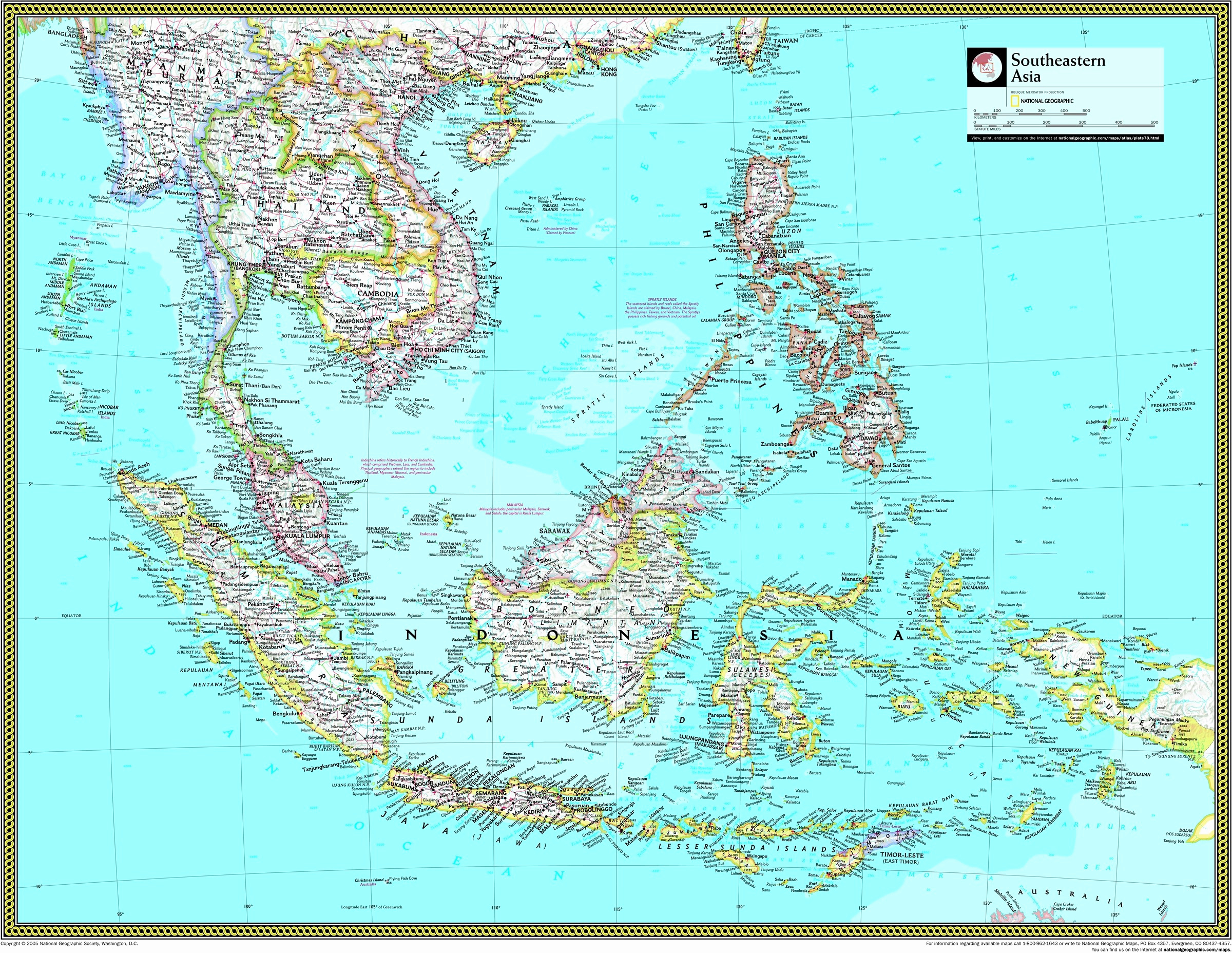} &
\includegraphics[width=1.5 cm, height=1.5 cm]{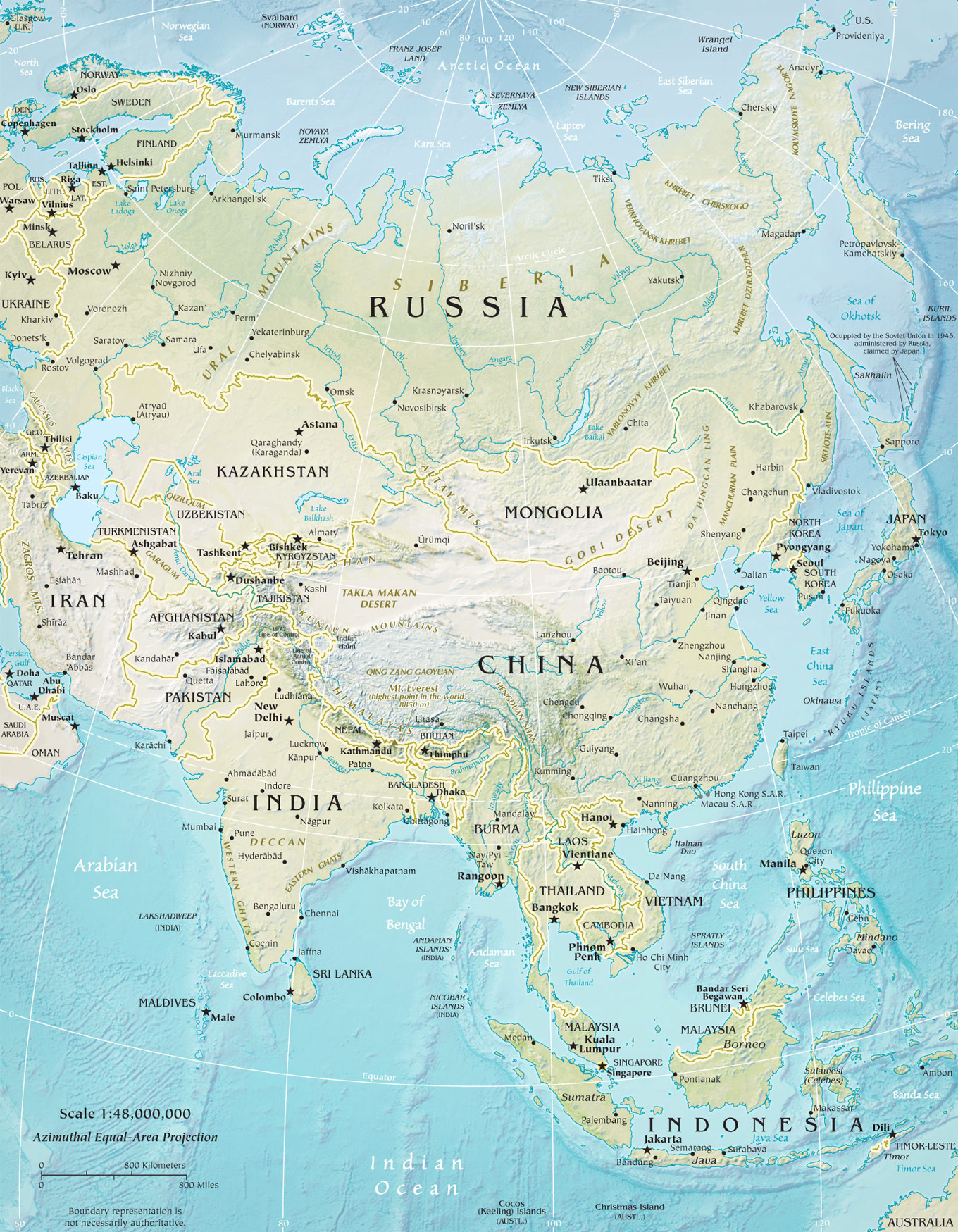}
\\

\text{m. Map - 1}  & \text{n. Map  - 2} & \text{o. Map  - 3} & \text{p. Map  - 4} 
\\

\end{tabular} 
\caption{Figures in different categories} 
\label{table:4} 
\end{figure}

\begin{figure} 
\scriptsize
\centering
\begin{tabular}{c c c c}

\includegraphics[width=1.5 cm, height=1.5 cm]{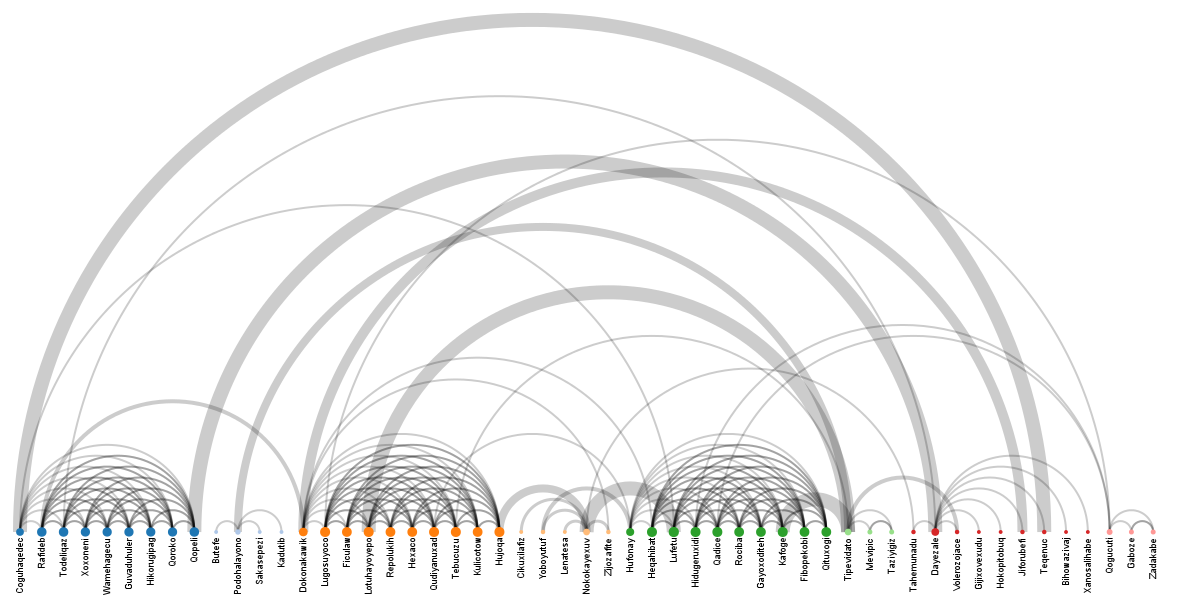} & 
\includegraphics[width=1.5 cm, height=1.5 cm]{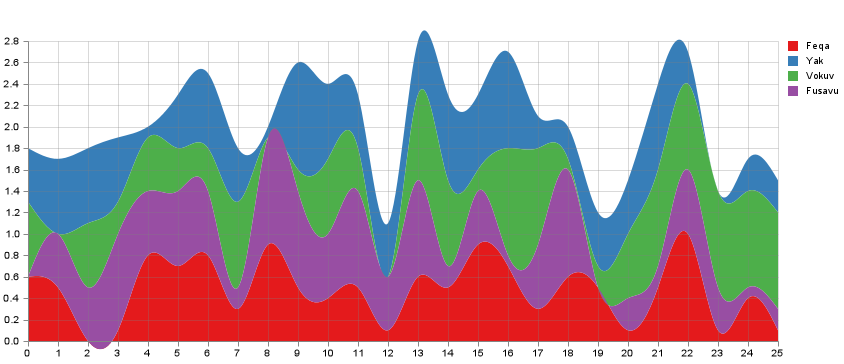} &
\includegraphics[width=1.5 cm, height=1.5 cm]{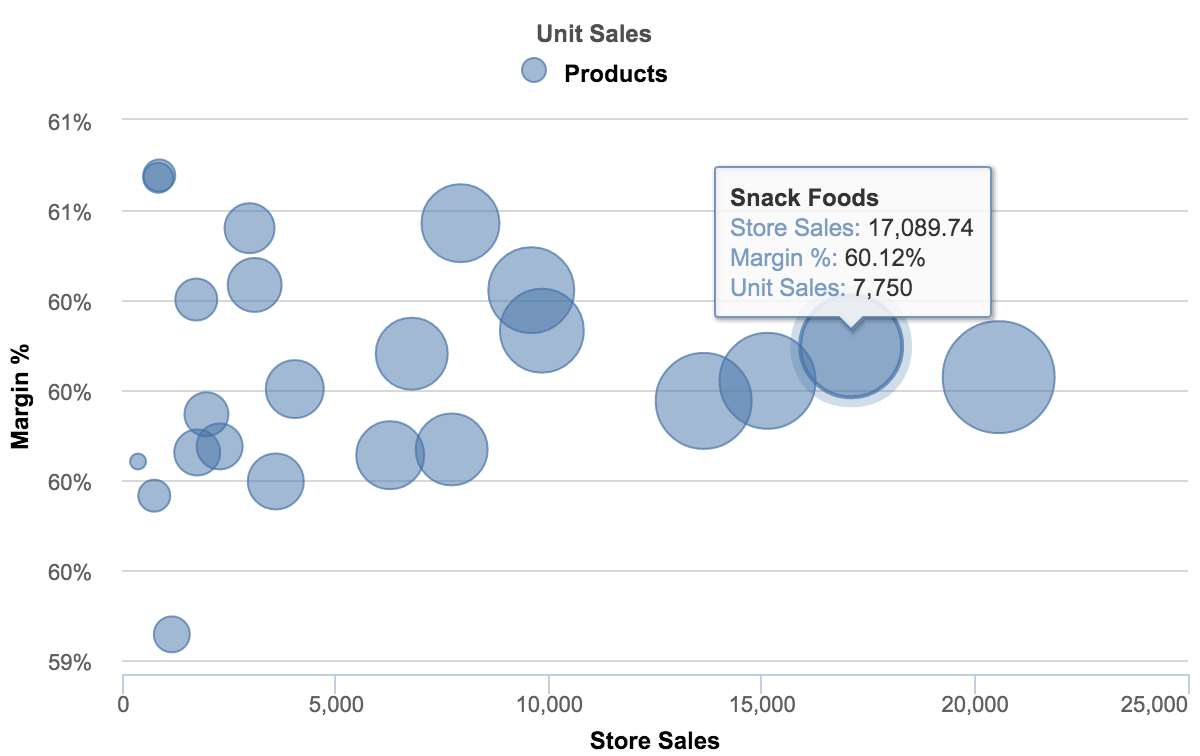} &
\includegraphics[width=1.5 cm, height=1.5 cm]{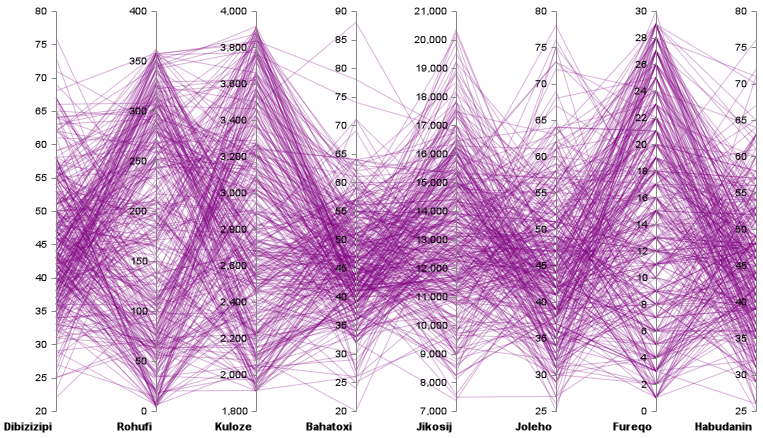}
\\

\text{a. Arc}  & \text{b. Area} & \text{c. Bubble} & \text{d. Parallel} \\
& & & \text{ Co-ordinate}
\\
\\

\includegraphics[width=1.5 cm, height=1.5 cm]{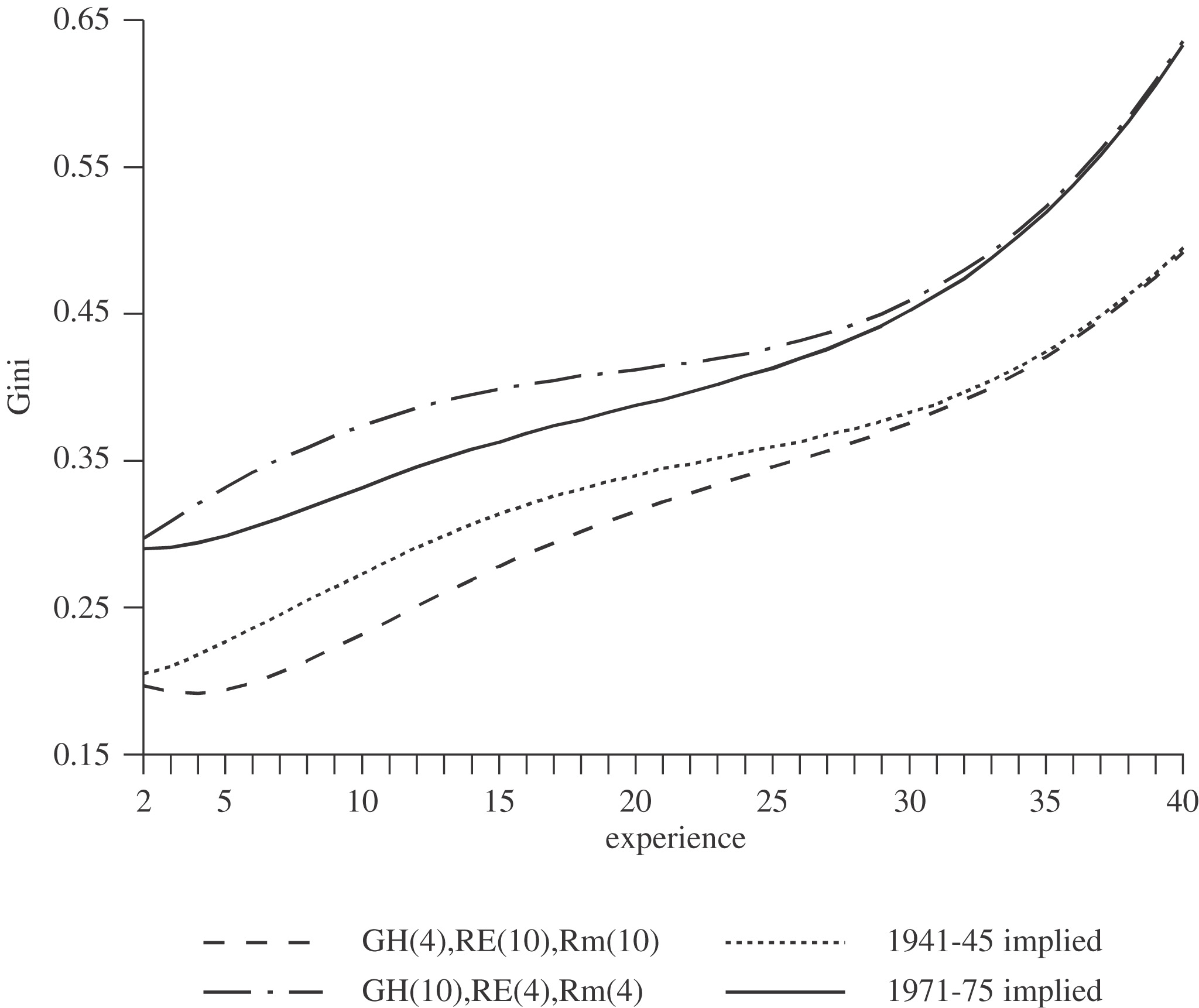} & 
\includegraphics[width=1.5 cm, height=1.5 cm]{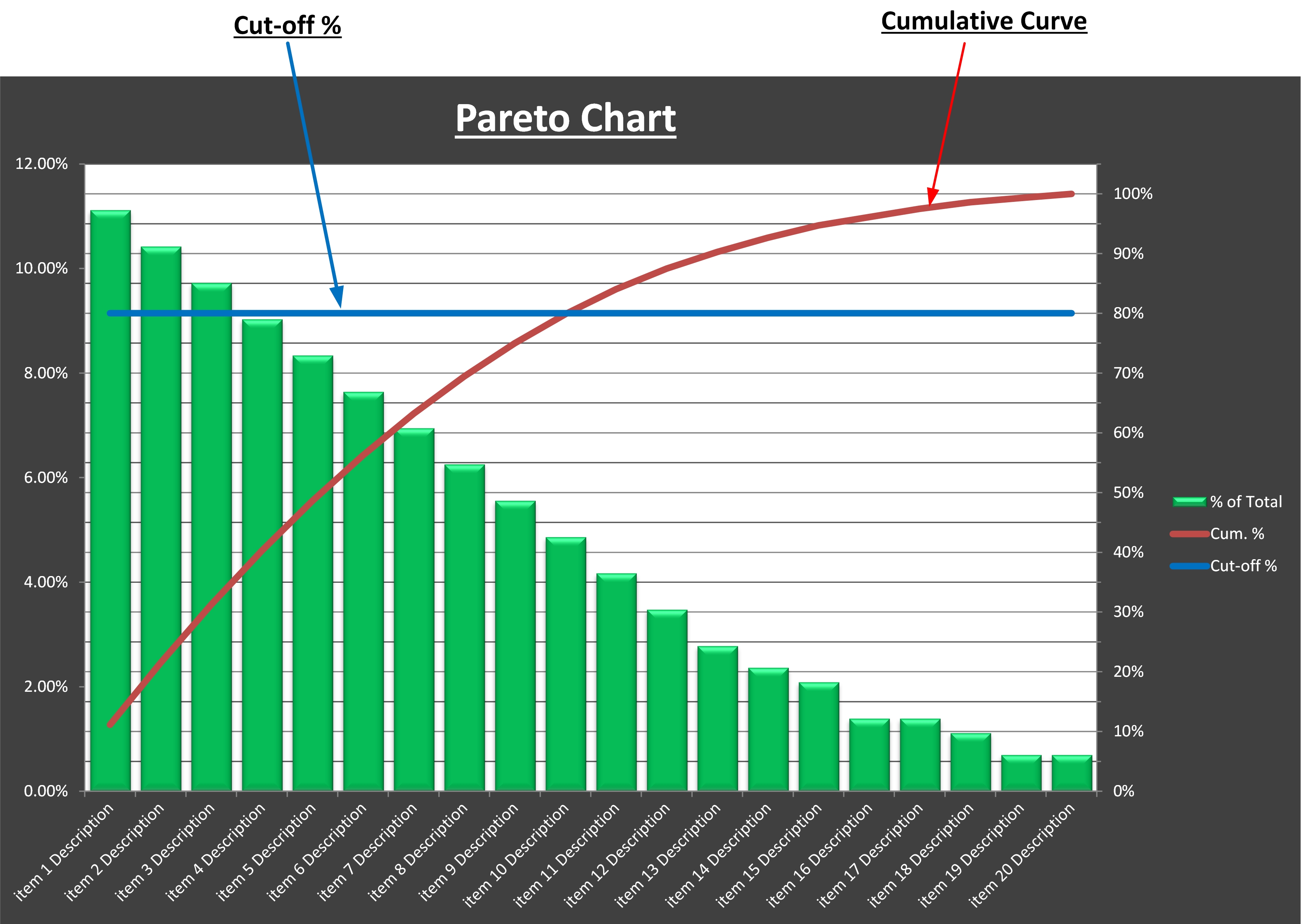} &
\includegraphics[width=1.5 cm, height=1.5 cm]{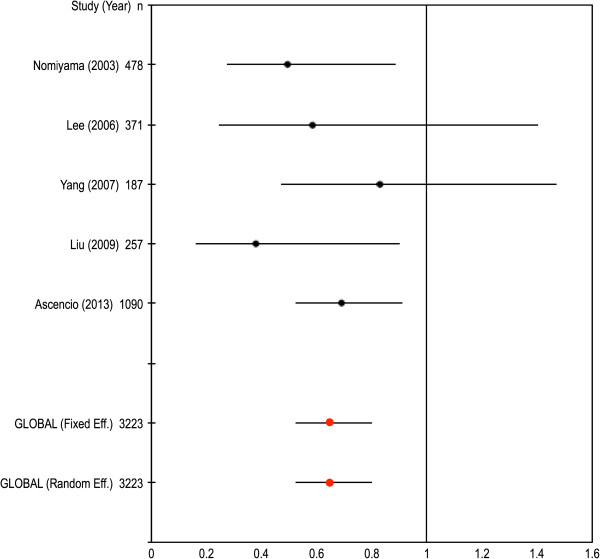} &
\includegraphics[width=1.5 cm, height=1.5 cm]{images/Plots/PMC_Horizontaol_Bar.jpg}
\\

\text{e. Line}  & \text{f. Pareto} & \text{g. Horizontal} & \text{h. Horizontal} \\
& & \text{Interval} & \text{Bar}
\\
\\

\includegraphics[width=1.5 cm, height=1.5 cm]{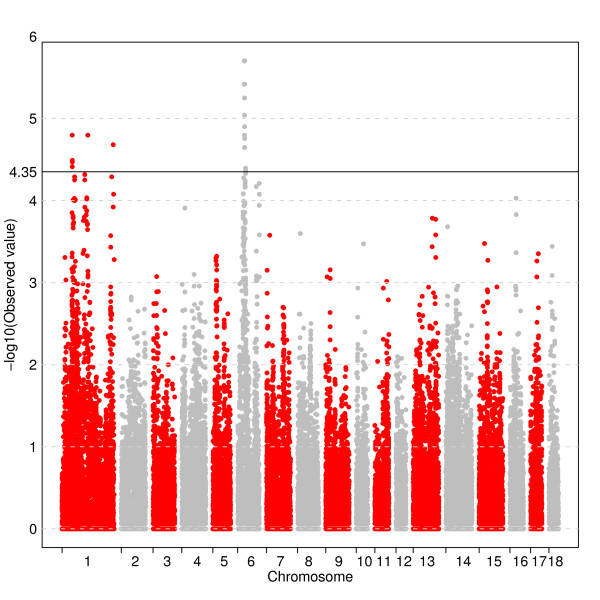} & 
\includegraphics[width=1.5 cm, height=1.5 cm]{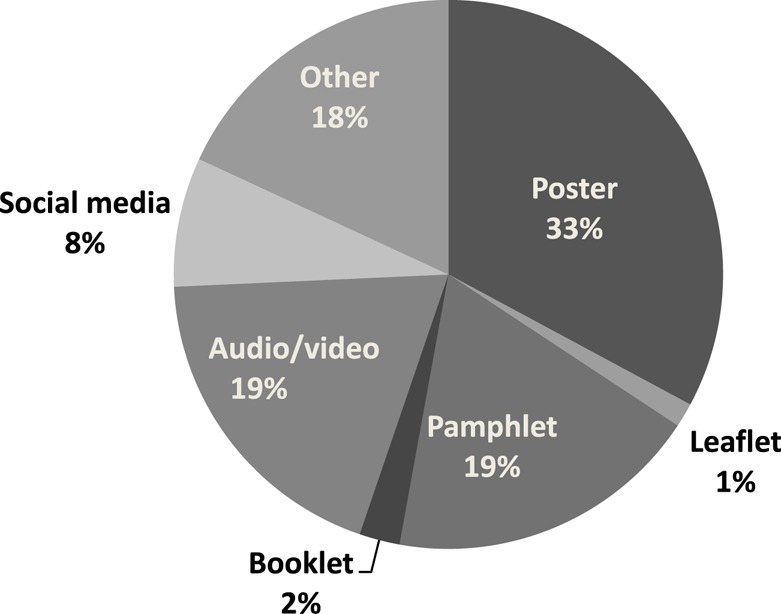} &
\includegraphics[width=1.5 cm, height=1.5 cm]{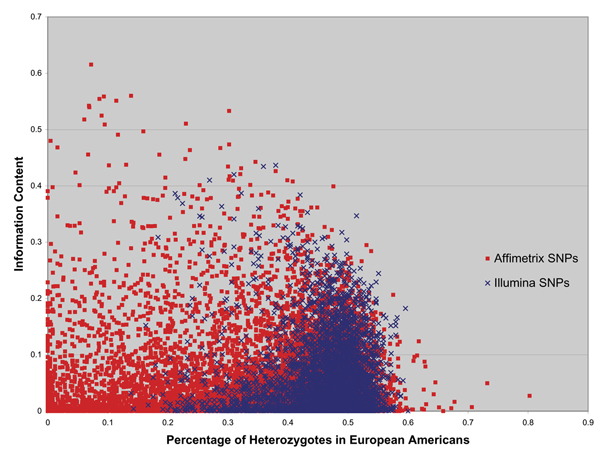} &
\includegraphics[width=1.5 cm, height=1.5 cm]{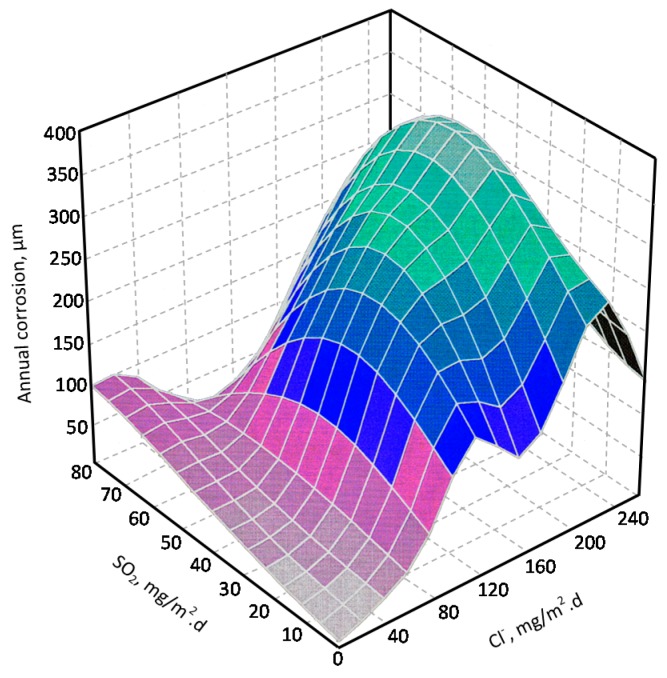}
\\
\text{i. Manhattan}  & \text{j. Pie} & \text{k. Scatter} & \text{l. Surface} 
\\
\\

\includegraphics[width=1.5 cm, height=1.5 cm]{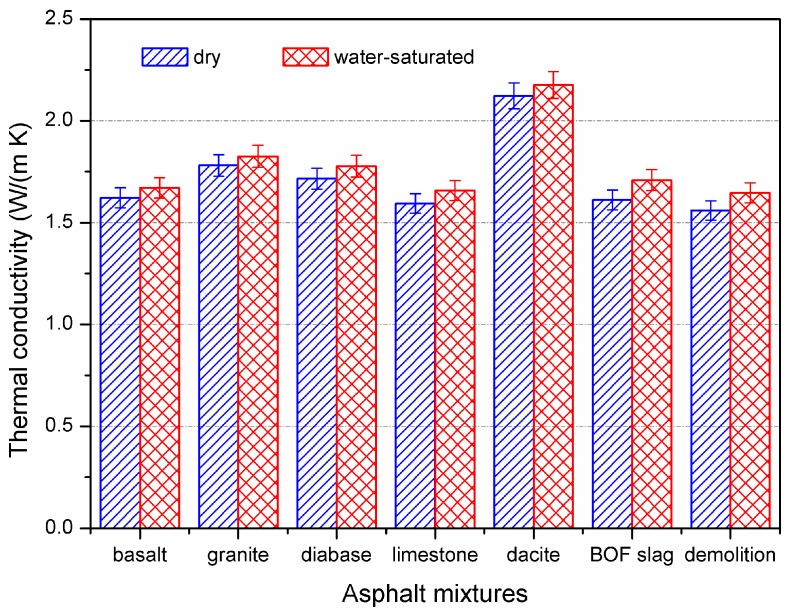} & 
\includegraphics[width=1.5 cm, height=1.5 cm]{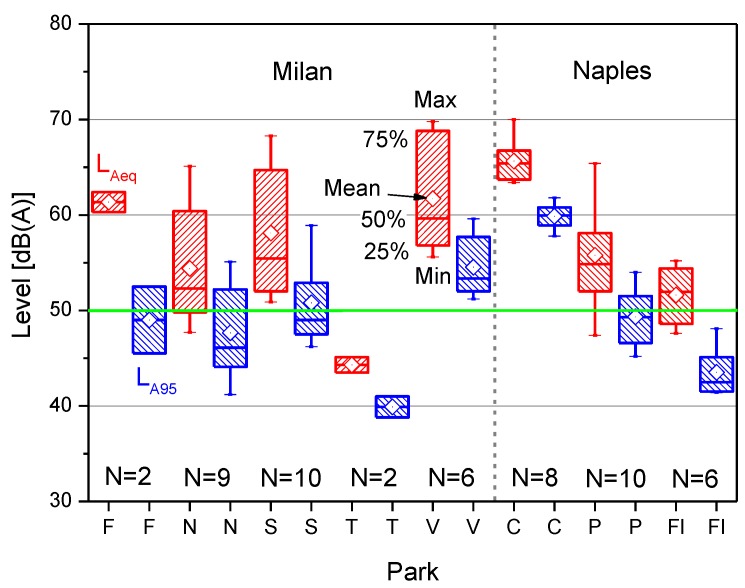} &
\includegraphics[width=1.5 cm, height=1.5 cm]{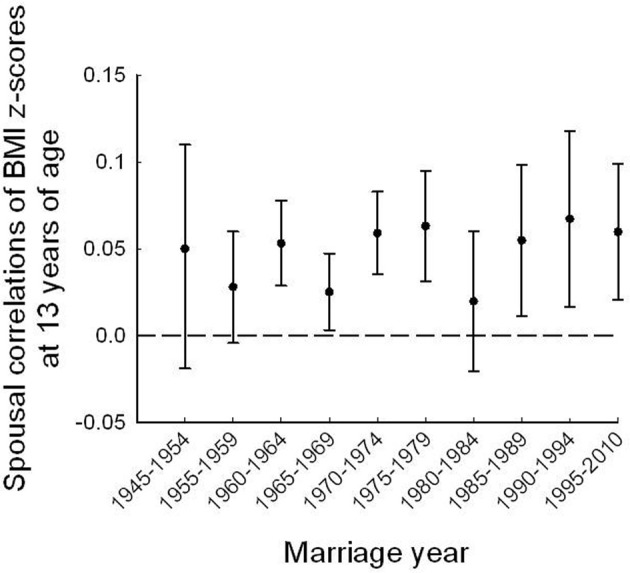} &
\includegraphics[width=1.5 cm, height=1.5 cm]{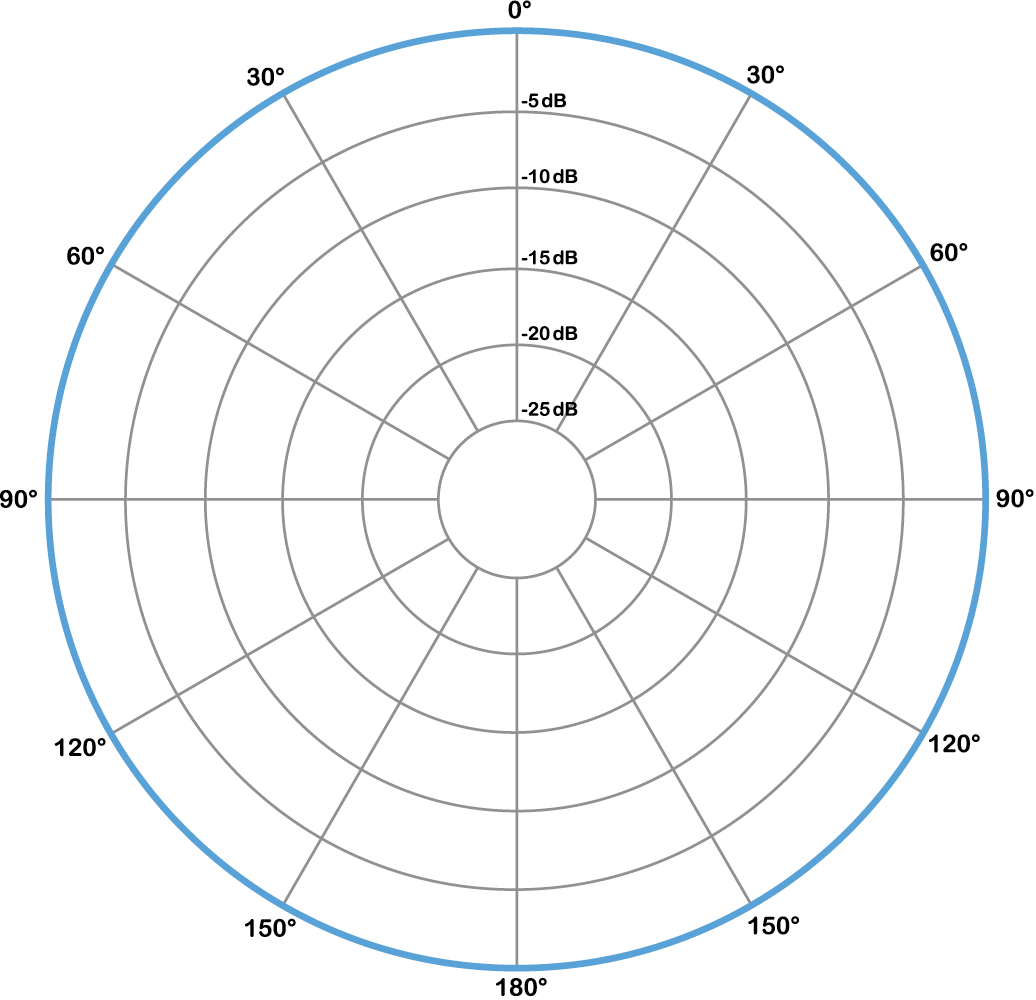}
\\
\text{m. Vertical Bar}  & \text{n. Box Plot} & \text{o. Vertical} & \text{p. Polar}\\
& & \text{Interval} & 
\\ 
\\

\includegraphics[width=1.5 cm, height=1.5 cm]{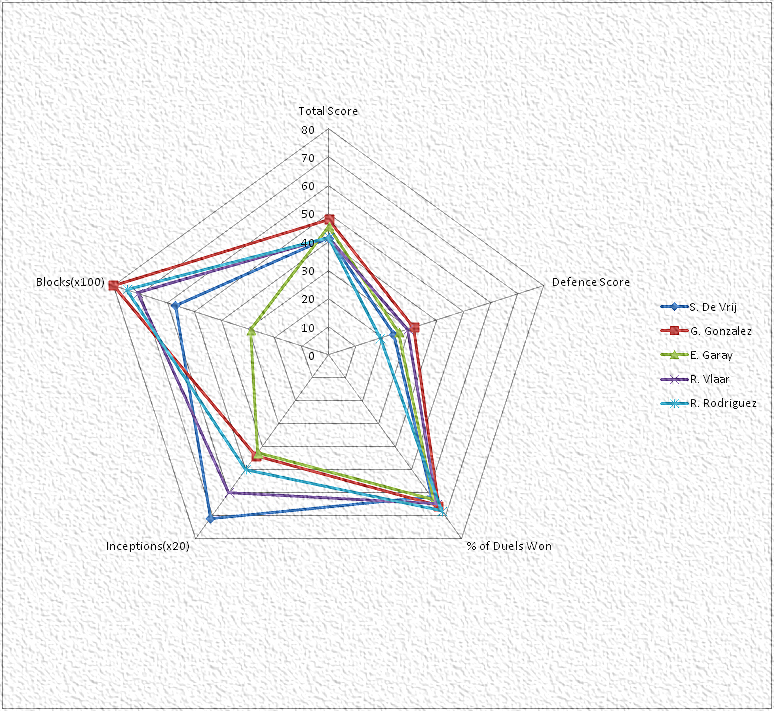} & 
\includegraphics[width=1.5 cm, height=1.5 cm]{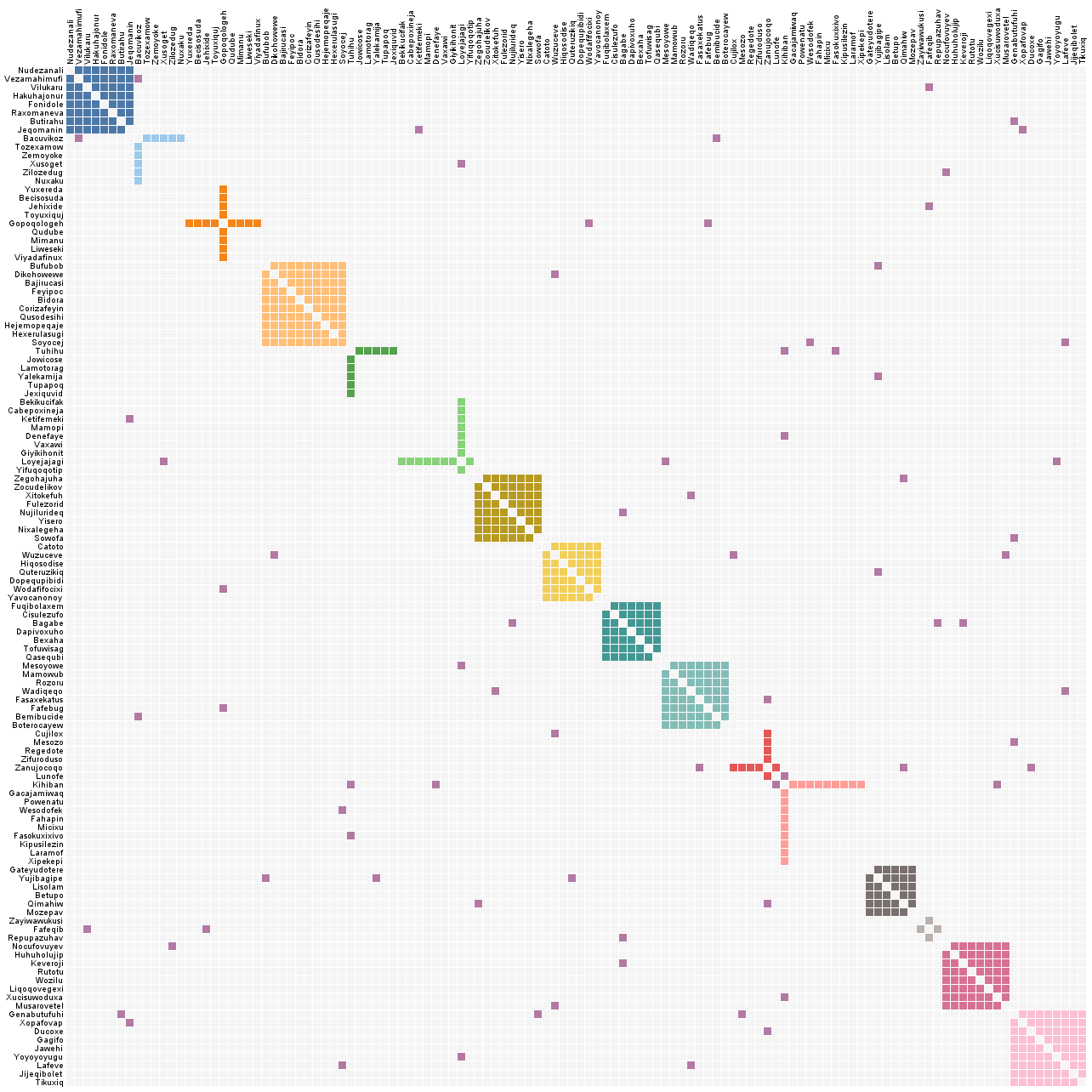} &
\includegraphics[width=1.5 cm, height=1.5 cm]{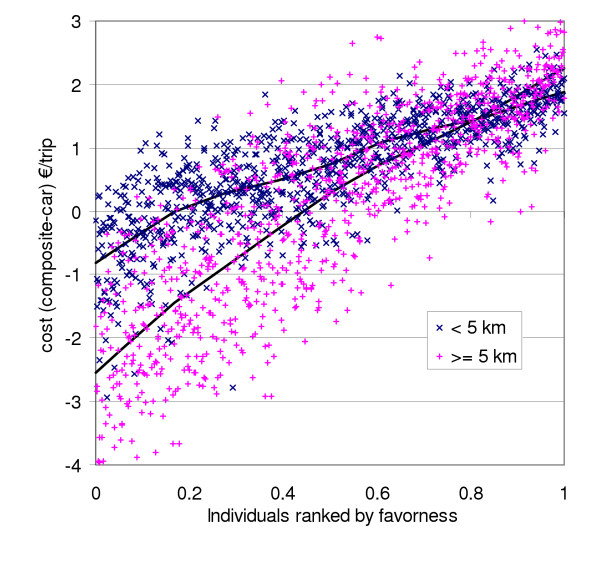} &
\includegraphics[width=1.5 cm, height=1.5 cm]{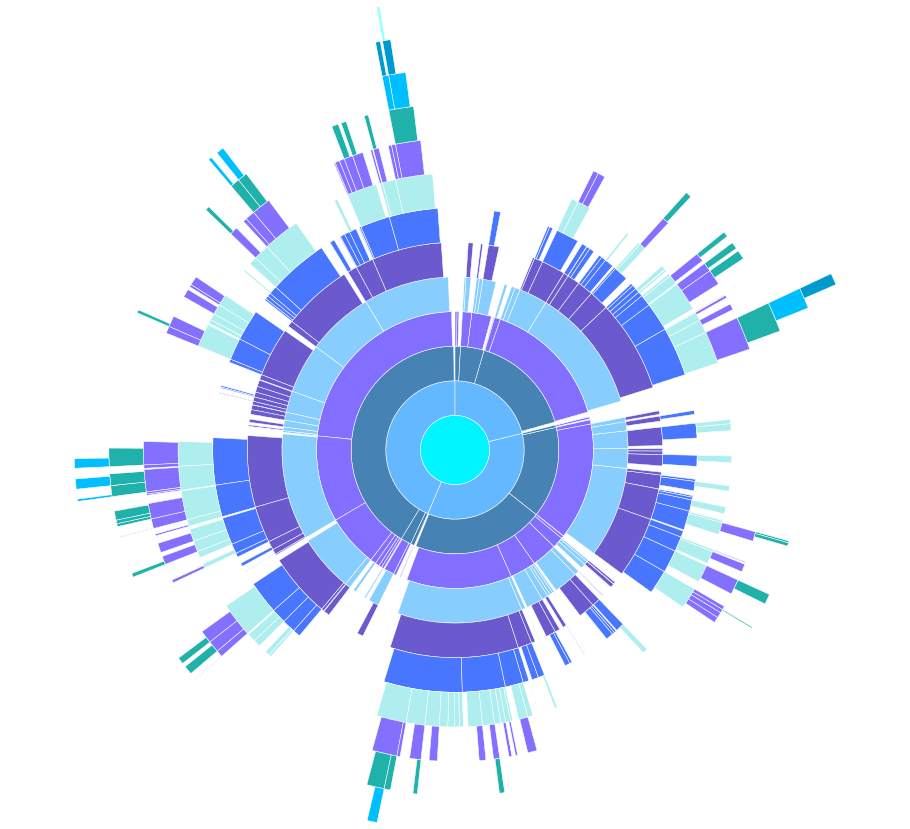}
\\
\text{q. Radar}  & \text{r. Reorderable} & \text{s. Scatter Line} & \text{t. Sunburst} \\
& \text{Matrix} & &
\\ 
\\
\includegraphics[width=1.5 cm, height=1.5 cm]{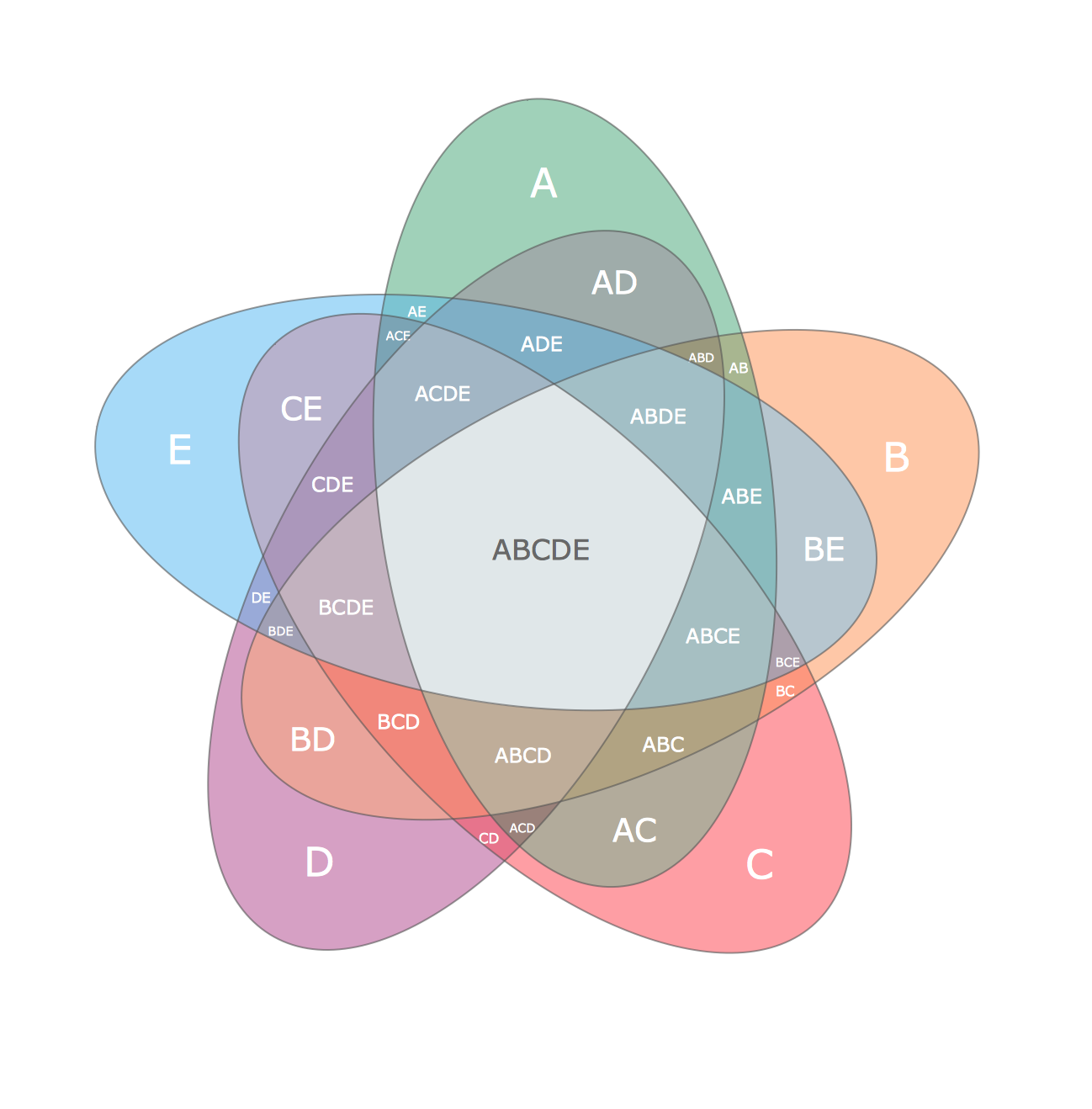} & 
\includegraphics[width=1.5 cm, height=1.5 cm]{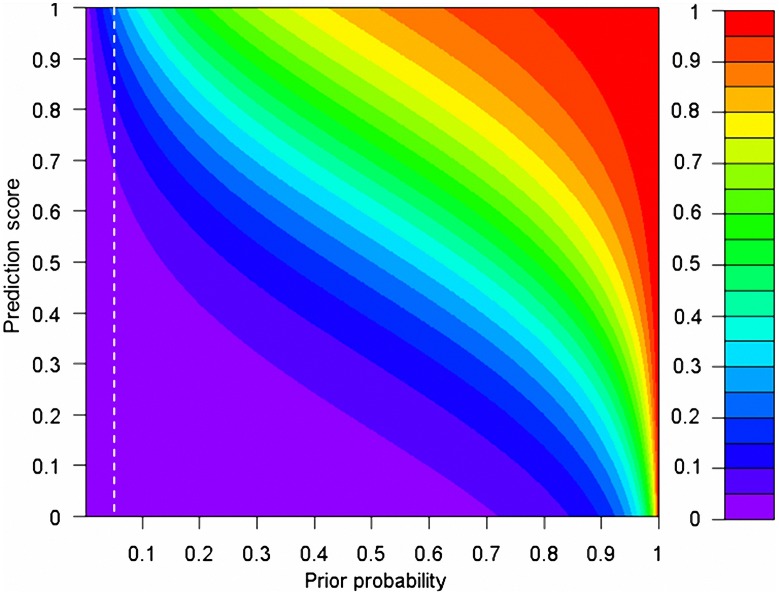} & &
\\
\text{u. Venn}  & \text{v. Heatmap} & & 
\\ 
\\
\end{tabular} 
\caption{Charts in different categories} 
\label{table:3} 
\end{figure}

\section{Future Directions}
Although there has been a significant increase in published articles on this classification problem, severe problems still need to be addressed. 
\subsection{\textbf{Lack of Benchmark Data set}}
The chart image classification problem has been extensively addressed in previous work. However, the high-level classification of charts from other types of figures needs a more state-of-the-art approach. ImageCLEF dataset includes a variety of figure types but is restricted to images in the medical domain. In addition to this, DocFigure and Slideimages have several different figure types. Still, there is a lack of state-of-the-art data sets to address the figure classification problem. Hence, there is a need for a dataset that includes a significant number of images and figure categories that would cover as many different figure types as possible.

\subsection{\textbf{Lack of Robust Model}}
Recent work makes some hard assumptions while addressing this problem. Most existing data sets contain a small number of real-figure images extracted from documents. This leads non-robust systems to fail when image samples contain intra-class dissimilarity or inter-class similarity. Including authentic figure images in the training phase could improve model performance. 

\subsection{\textbf{Inclusion of Noise}}
Most of the work in the existing literature ignores the effect of noise. The presence of different types of noise, such as background grids, low image quality, composite charts, and the presence of multiple components along with figures, leads to poor performance for models that perform exceptionally on noiseless data\cite{thiyam_challenges_2021}.   

So, there is a need for a robust deep-learning model to cover all the shortcomings mentioned above.

\section{Conclusion}
Figure classification is challenging due to the variety of figures present, the similarity between different figure types, and the noise in the figure images. Techniques used for figure classification have evolved remarkably. Earlier methods focused on manual feature extraction and providing the feature vectors to the different classifiers. Recent approaches, however, use more specific features corresponding to specific figure types more efficiently using deep learning models. Though the performance of these techniques is good, they are not robust enough to handle noisy and real figure image data. In this survey, various methods used for figure classification were discussed, along with the publicly available data sets. Also, some pointers are provided for the shortcomings in the current works.

 \bibliographystyle{ieeetr}

\bibliography{ad}

\begin{thebibliography}{10}

\bibitem{ATM36944}
L.~Cai, J.~Gao, and D.~Zhao, ``A review of the application of deep learning in
  medical image classification and segmentation,'' {\em Annals of Translational
  Medicine}, vol.~8, no.~11, 2020.

\bibitem{RaniBR2018ClassificationOV}
S.~Rani.B.R, ``Classification of vehicles using image processing techniques,''
  {\em International journal of engineering research and technology}, vol.~3,
  2018.

\bibitem{LIU2021223}
L.~Liu, Z.~Wang, T.~Qiu, Q.~Chen, Y.~Lu, and C.~Y. Suen, ``Document image
  classification: Progress over two decades,'' {\em Neurocomputing}, vol.~453,
  pp.~223--240, 2021.

\bibitem{Kumar2011SurveyOT}
M.~Kumar, M.~Kamble, S.~Pawar, P.~Patil, and N.~Bonde, ``Survey on techniques
  for plant leaf classification,'' 2011.

\bibitem{davila_chart_2021}
K.~Davila, S.~Setlur, D.~Doermann, B.~U. Kota, and V.~Govindaraju, ``Chart
  {Mining}: {A} {Survey} of {Methods} for {Automated} {Chart} {Analysis},''
  {\em IEEE Transactions on Pattern Analysis and Machine Intelligence},
  vol.~43, pp.~3799--3819, Nov. 2021.
\newblock Conference Name: IEEE Transactions on Pattern Analysis and Machine
  Intelligence.

\bibitem{siegel_figureseer_2016}
N.~Siegel, Z.~Horvitz, R.~Levin, S.~Divvala, and A.~Farhadi, ``{FigureSeer}:
  {Parsing} {Result}-{Figures} in {Research} {Papers},'' in {\em {ECCV}}, 2016.

\bibitem{lee_viziometrics_2018}
P.-S. Lee, J.~D. West, and B.~Howe, ``Viziometrics: {Analyzing} {Visual}
  {Information} in the {Scientific} {Literature},'' {\em IEEE Transactions on
  Big Data}, vol.~4, pp.~117--129, Mar. 2018.
\newblock Conference Name: IEEE Transactions on Big Data.

\bibitem{morris_slideimages_2020}
D.~Morris, E.~Müller-Budack, and R.~Ewerth, ``{SlideImages}: {A} {Dataset} for
  {Educational} {Image} {Classification},'' {\em arXiv:2001.06823 [cs]}, Jan.
  2020.
\newblock arXiv: 2001.06823.

\bibitem{kv_docfigure_2019}
J.~kv, A.~Mondal, and C.~Jawahar, ``{DocFigure}: {A} {Dataset} for {Scientific}
  {Document} {Figure} {Classification},'' pp.~74--79, Sept. 2019.

\bibitem{andrearczyk_deep_2018}
V.~Andrearczyk and H.~Müller, ``Deep {Multimodal} {Classification} of {Image}
  {Types} in {Biomedical} {Journal} {Figures},'' in {\em {CLEF}}, 2018.

\bibitem{almakky_stacked_2018}
I.~Almakky, V.~Palade, Y.-L. Hedley, and J.~Yang, ``A stacked deep autoencoder
  model for biomedical figure classification,'' in {\em 2018 {IEEE} 15th
  {International} {Symposium} on {Biomedical} {Imaging} ({ISBI} 2018)},
  pp.~1134--1138, Apr. 2018.
\newblock ISSN: 1945-8452.

\bibitem{lu_automated_2009}
X.~Lu, S.~Kataria, W.~J. Brouwer, J.~Z. Wang, P.~Mitra, and C.~L. Giles,
  ``Automated analysis of images in documents for intelligent document
  search,'' {\em International Journal on Document Analysis and Recognition
  (IJDAR)}, vol.~12, pp.~65--81, July 2009.

\bibitem{cheng_graphical_2013}
B.~Cheng, R.~Stanley, S.~Antani, and G.~Thoma, ``Graphical {Figure}
  {Classification} {Using} {Data} {Fusion} for {Integrating} {Text} and {Image}
  {Features},'' {\em Proceedings of the International Conference on Document
  Analysis and Recognition, ICDAR}, Aug. 2013.

\bibitem{lagopoulos_classifying_2019}
A.~Lagopoulos, N.~Kapraras, V.~Amanatiadis, A.~Fachantidis, and G.~Tsoumakas,
  ``Classifying {Biomedical} {Figures} by {Modality} via {Multi}-{Label}
  {Learning},'' {\em IEEE Journal of Biomedical and Health Informatics},
  vol.~23, pp.~2230--2237, Nov. 2019.
\newblock Conference Name: IEEE Journal of Biomedical and Health Informatics.

\bibitem{giannakopoulos_visual-based_2015}
T.~Giannakopoulos, Y.~Foufoulas, E.~Stamatogiannakis, H.~Dimitropoulos,
  N.~Manola, and Y.~Ioannidis, ``Visual-{Based} {Classification} of {Figures}
  from {Scientific} {Literature},'' pp.~1059--1060, May 2015.

\bibitem{hashmi_current_2021}
K.~A. Hashmi, M.~Liwicki, D.~Stricker, M.~A. Afzal, M.~A. Afzal, and M.~Z.
  Afzal, ``Current {Status} and {Performance} {Analysis} of {Table}
  {Recognition} in {Document} {Images} with {Deep} {Neural} {Networks},'' {\em
  arXiv:2104.14272 [cs]}, May 2021.
\newblock arXiv: 2104.14272.

\bibitem{savva_revision_2011}
M.~Savva, N.~Kong, A.~Chhajta, L.~Fei-Fei, M.~Agrawala, and J.~Heer,
  ``{ReVision}: automated classification, analysis and redesign of chart
  images,'' in {\em Proceedings of the 24th annual {ACM} symposium on {User}
  interface software and technology}, {UIST} '11, (New York, NY, USA),
  pp.~393--402, Association for Computing Machinery, Oct. 2011.

\bibitem{gao_view_2012}
J.~Gao, Y.~Zhou, and K.~E. Barner, ``View: {Visual} {Information} {Extraction}
  {Widget} for improving chart images accessibility,'' in {\em 2012 19th {IEEE}
  {International} {Conference} on {Image} {Processing}}, pp.~2865--2868, Sept.
  2012.
\newblock ISSN: 2381-8549.

\bibitem{karthikeyani_machine_2012}
V.~Karthikeyani and S.~Nagarajan, ``Machine {Learning} {Classification}
  {Algorithms} to {Recognize} {Chart} {Types} in {Portable} {Document} {Format}
  ({PDF}) {Files},'' {\em International Journal of Computer Applications},
  vol.~39, pp.~1--5, Feb. 2012.

\bibitem{liu_chart_2015}
X.~Liu, B.~Tang, Z.~Wang, X.~Xu, S.~Pu, D.~Tao, and M.~Song, ``Chart
  classification by combining deep convolutional networks and deep belief
  networks,'' in {\em 2015 13th {International} {Conference} on {Document}
  {Analysis} and {Recognition} ({ICDAR})}, pp.~801--805, Aug. 2015.

\bibitem{amara_convolutional_2017}
J.~Amara, P.~Kaur, M.~Owonibi, and B.~Bouaziz, ``Convolutional {Neural}
  {Network} {Based} {Chart} {Image} {Classification},'' May 2017.

\bibitem{jung_chartsense_2017}
D.~Jung, W.~Kim, H.~Song, J.~Hwang, B.~Lee, B.~H. Kim, and J.~Seo,
  ``{ChartSense}: {Interactive} {Data} {Extraction} from {Chart} {Images},''
  {\em CHI}, 2017.

\bibitem{sanderson2019imageclef}
M.~Sanderson and P.~Clough, ``Imageclef—the clef cross language image
  retrieval track| imageclef/lifeclef—multimedia retrieval in clef,'' 2019.

\bibitem{balaji_chart-text_2018}
A.~Balaji, T.~Ramanathan, and V.~Sonathi, ``Chart-{Text}: {A} {Fully}
  {Automated} {Chart} {Image} {Descriptor},'' Dec. 2018.
\newblock arXiv:1812.10636 [cs].

\bibitem{chagas_evaluation_2018}
P.~Chagas, R.~Akiyama, A.~Meiguins, C.~Santos, F.~Saraiva, B.~Meiguins, and
  J.~Morais, ``Evaluation of {Convolutional} {Neural} {Network} {Architectures}
  for {Chart} {Image} {Classification},'' in {\em 2018 {International} {Joint}
  {Conference} on {Neural} {Networks} ({IJCNN})}, pp.~1--8, July 2018.
\newblock ISSN: 2161-4407.

\bibitem{dai_chart_2018}
W.~Dai, M.~Wang, Z.~Niu, and J.~Zhang, ``Chart decoder: {Generating} textual
  and numeric information from chart images automatically,'' {\em Journal of
  Visual Languages \& Computing}, vol.~48, pp.~101--109, Oct. 2018.

\bibitem{liu_data_2019}
X.~Liu, D.~Klabjan, and P.~NBless, ``Data {Extraction} from {Charts} via
  {Single} {Deep} {Neural} {Network},'' June 2019.
\newblock arXiv:1906.11906 [cs].

\bibitem{davila_icdar_2019}
K.~Davila, B.~U. Kota, S.~Setlur, V.~Govindaraju, C.~Tensmeyer, S.~Shekhar, and
  R.~Chaudhry, ``{ICDAR} 2019 {Competition} on {Harvesting} {Raw} {Tables} from
  {Infographics} ({CHART}-{Infographics}),'' in {\em 2019 {International}
  {Conference} on {Document} {Analysis} and {Recognition} ({ICDAR})}, (Sydney,
  Australia), pp.~1594--1599, IEEE, Sept. 2019.

\bibitem{bajic_data_2020}
F.~Bajić, J.~Job, and K.~Nenadić, ``Data {Visualization} {Classification}
  {Using} {Simple} {Convolutional} {Neural} {Network} {Model},'' {\em
  International Journal of Electrical and Computer Engineering Systems
  (IJECES)}, vol.~11, no.~1, pp.~43--51, 2020.

\bibitem{araujo_real-world_2020}
T.~Araújo, P.~Chagas, J.~Alves, C.~Santos, B.~Sousa~Santos, and
  B.~Serique~Meiguins, ``A {Real}-{World} {Approach} on the {Problem} of
  {Chart} {Recognition} {Using} {Classification}, {Detection} and {Perspective}
  {Correction},'' {\em Sensors}, vol.~20, p.~4370, Jan. 2020.
\newblock Number: 16 Publisher: Multidisciplinary Digital Publishing Institute.

\bibitem{luo_chartocr_2021}
J.~Luo, Z.~Li, J.~Wang, and C.-Y. Lin, ``{ChartOCR}: {Data} {Extraction} from
  {Charts} {Images} via a {Deep} {Hybrid} {Framework},'' in {\em 2021 {IEEE}
  {Winter} {Conference} on {Applications} of {Computer} {Vision} ({WACV})},
  (Waikoloa, HI, USA), pp.~1916--1924, IEEE, Jan. 2021.

\bibitem{davila_icpr_2021}
K.~Davila, C.~Tensmeyer, S.~Shekhar, H.~Singh, S.~Setlur, and V.~Govindaraju,
  ``{ICPR} 2020 - {Competition} on {Harvesting} {Raw} {Tables} from
  {Infographics},'' pp.~361--380, Feb. 2021.

\bibitem{thiyam_challenges_2021}
J.~Thiyam, S.~R. Singh, and P.~K. Bora, ``Challenges in chart image
  classification: a comparative study of different deep learning methods,'' in
  {\em Proceedings of the 21st {ACM} {Symposium} on {Document} {Engineering}},
  {DocEng} '21, (New York, NY, USA), pp.~1--4, Association for Computing
  Machinery, Aug. 2021.

\end{thebibliography}
\end{document}